\documentclass[aps,prb,reprint,groupedaddress,superscriptaddress,floatfix,longbibliography]{revtex4-2}

\usepackage[T1]{fontenc}
\usepackage{newtxtext}
\usepackage{newtxmath}

\usepackage{titlesec}
\usepackage{appendix}
\usepackage{amsmath}
\usepackage{mathtools}
\usepackage{amsfonts}
\usepackage{bbold}
\usepackage{physics}
\usepackage{color}
\usepackage{xcolor}
\usepackage[colorlinks,citecolor=blue,linkcolor=blue,urlcolor=blue]{hyperref}
\usepackage{hyperref}
\usepackage{graphicx}
\usepackage[all]{hypcap}

\usepackage{enumitem}

\usepackage{tabularx}
\usepackage{makecell}

\begin{document}

  \title{Exact Solvability and Integrability Signatures in a Periodically Driven Infinite-Range  $p$-Spin \\ Kicked Top Model}

\author{Harshit Sharma}
\email{harshitsharma2796@gmail.com}
\affiliation{Department of Physics, Visvesvaraya  National Institute of Technology, Nagpur 440010, India}

\author{Sashmita Rout}
\email{sashmitaa111@gmail.com}
\affiliation{Department of Physics, Visvesvaraya  National Institute of Technology, Nagpur 440010, India}

\author{Avadhut V. Purohit}
\email{avdhoot.purohit@gmail.com}
\affiliation{Department of Physics, Visvesvaraya  National Institute of Technology, Nagpur 440010, India}

\author{Udaysinh T. Bhosale}
\email{udaysinhbhosale@phy.vnit.ac.in}
\affiliation{Department of Physics, Visvesvaraya  National Institute of Technology, Nagpur 440010, India}

\date{\today}

\begin{abstract}
We study the signatures of quantum integrability (QI) in a $p$-spin kicked top model with infinite-range Ising interactions, subjected to periodic pulses of an external magnetic field, with the precession angle fixed at $\alpha=m\pi$. We analytically compute the unitary operator, its eigensystem, time evolution, single-qubit reduced density matrix, and entanglement dynamics for arbitrary initial states. We show that the spectrum of the unitary operator  exhibits pronounced clustering of eigenphases in some regions, leading to strongly non-uniform spectral features with $p$. The entanglement measures and unitary operator dynamics exhibit periodic behavior, while the eigenvalue spectrum shows strong degeneracies when the kicking strength $k'$ is a rational multiple of $\pi$ for even $p$.
For odd $p$, both the periodic behavior and spectral degeneracies persist for all values of $k'$. This periodic behavior extends to the strong-kicking regime ($k'=J  j^{p-1}\pi$), where for rational values of $J$ the dynamics remain periodic. For any $k'$ and $p$ the spectral statistics shows Poissonian behavior and  the average adjacent gap ratio $\langle r \rangle = 2\ln 2 - 1$, consistent with integrable behaviour. The spectral form factor further supports this nature. Additionally, we compute  the ratio of average eigenstate entanglement entropy to its maximum value ($\langle S \rangle /S_{Max}$) and find that it remains significantly far below from $1$ in the limit $N\rightarrow \infty$. These results collectively establish robust signatures of QI across all $k’, p,$ and system sizes. We also discuss possible experimental realizations of the model.
\end{abstract}
\maketitle
\section{ Introduction}
The integrable models play a prominent role in various areas of physics, such as condensed matter, quantum computing, statistical mechanics, quantum groups and Yangians, AdS/CFT, the Hubbard model, sigma models, and many others \cite{shastry1986exact,van2022preparing,delduc2019integrable,batchelor2016yang,beisert2012review,retore2022introduction,calabrese2020entanglement}. They are extensively studied due to their high degree of symmetry, which imposes strong constraints on their dynamics and enable exact solvability. Integrable models come in various forms, with the two main classes being continuum  models and lattice models. The continuum model includes conformal field theories \cite{ketov1995conformal,kazakov2018biscalar}, massive relativistic models like the sine-Gordon \cite{cuevas2014sine},  and nonlinear sigma models \cite{kurkccuouglu2008noncommutative,fateev2018integrability}, and non-relativistic models such as the Lieb–Liniger model \cite{mailoud2021spectrum}. In contrast, the lattice model includes interacting fermion and boson theories \cite{reis2022resurgence,guan2022new}, electronic models such as the Hubbard model \cite{shastry1986exact}, and the spin chains like the Heisenberg spin chain \cite{andrei1984heisenberg}, XY chains \cite{its2005entanglement}, and the transverse-field Ising interacting models \cite{mishra2015protocol,doikou2010introduction,pal2018entangling,sharma2024exactly,sharma2024signatures}.

The spin chain model is extensively studied mainly in two classes: the nearest-neighbor interaction and the long-range interaction. The long-range interactions  decay as power law $1/r^{\gamma}$, as a function of distance $r$ \cite{campa2014physics,dauxois2002dynamics,defenu2023long}. By tuning the exponent $\gamma$, the long-range interactions fall into several categories, with particular interest in the infinite-range interaction ($\gamma=0$) \cite{britton2012engineered,
jurcevic2014quasiparticle,douglas2015quantum,gil2016nonequilibrium,Marino2019,sauerwein2023engineering,liu2024signature,vzunkovivc2024mean,qi2023surprises}. These interactions play a crucial role in various quantum technology applications like quantum computing \cite{inoue2015infinite,lewis2023ion}, quantum heat engine \cite{solfanelli2023quantum}, ion trap \cite{gambetta2020long} etc. The entanglement properties of the system can be analyzed by operating the time evolution unitary operator on the initial unentangled states. Various entanglement measures have been investigated in numerous interacting spin models \cite{latorre2009short,pal2018entangling,kumari2022eigenstate,sharma2024exactly}. Since entanglement generation is fundamentally governed by the nature of interactions between constituent spins, understanding the structure and behaviour of these interactions becomes essential. In most physical systems,  spin interactions are predominantly pairwise ($p=2$). However, higher-order ($p>2$) spin interactions naturally arise in several contexts, including high-energy and nuclear physics, as well as quantum circuits, algorithms in quantum chemistry, error correction codes and other applications \cite{banuls2020simulating,hauke2013quantum,hempel2018quantum,paetznick2013universal,kitaev2003fault,arias2021high,marvian2022restrictions}. From a fundamental perspective, $p$-body interactions with $p>2$ exhibit qualitatively different behavior compared to pairwise interactions, leading to many-body correlations, enhanced complexity, and nontrivial dynamical features \cite{munoz2021nonlinear}. It can be seen across multiple areas of physics, making the study of higher-body interactions a central research direction \cite{efimov1970energy,kraemer2006evidence}. Such interactions are also relevant in diverse quantum platforms, including neural networks \cite{weimer2010rydberg,levine2019parallel,xing2021realization,naidon2017efimov}, spin glasses \cite{kirkpatrick1987p,bardina2004p,Munoz-Arias2020,GARDNER1985747}, tensor networks \cite{lanthier2025tensor}, superconducting systems \cite{khazali2020fast,kim2022high,menke2022demonstration}, and trapped ions \cite{cirac1995quantum,katz2023demonstration}. Moreover, models with $p>2$ interactions often exhibit rugged energy landscapes, which makes them important benchmarks for both classical and quantum optimization algorithms \cite{kanao2023simulated,jorg2010first,bellitti2021entropic}.

Linear systems are typically integrable, whereas the inclusion of nonlinear interactions often promotes the onset of chaos. However, nonlinearity alone does not guarantee chaos. Various nonlinear systems can still exhibit integrability, which is often determined by the presence of underlying symmetries in specific parameter regimes \cite{sharma2024exactly,sharma2024signatures,sharma2025exact}. Indeed, several well-known integrable models are characterized by pairwise ($p=2$) interactions, including both nearest-neighbor and long-range interactions \cite{kumari2022eigenstate,sharma2024exactly,sharma2024signatures}. If we introduce stronger nonlinearities ($p>2$), then the natural  question arises: do such highly nonlinear systems inevitably lead to chaos for any parameter domain, or can they still exhibit integrable behavior? In this work, we investigate the signatures of integrability in models with $p$-spin interactions in the $\gamma=0$ regime. Remarkably, despite the presence of strong nonlinear terms, the system can remain integrable for specific parameter choices. This provides a natural platform to explore the interplay between nonlinearity and integrability in quantum many-body systems.

 To systematically investigate this interplay, it is essential to first clarify the notion of integrability in both classical and quantum contexts \cite{caux2011remarks}. In classical mechanics, a system is integrable if it satisfies the Liouville-Arnold theorem, which states that a system in a $2n$-dimensional phase space, there must exist $n$ independent conserved quantities in involution. \cite{babelon2003introduction,retore2022introduction,torrielli2016classical}. However,  classically integrable systems do not necessarily remain integrable in their quantum domain, even if they have many conserved quantities \cite{razavy1986does,hietarinta1982quantum}.
Various techniques have been developed to study classical integrable systems \cite{torrielli2016classical}. It is worth noting that there is no universally accepted definition of quantum integrability. It is commonly characterized by the existence of an extensive set of conserved quantities and, in many cases, by the exact solvability of the model \cite{thacker1981exact,owusu2008link,doikou2010introduction,gubin2012quantum,yuzbashyan2013quantum,
gombor2021integrable,tang2023integrability,vernier2024strong,claeys2018integrable,gritsev2017integrable}. This can be achieved through methods such as the Yang-Baxter equation \cite{wadati1993quantum,doikou2010introduction,lambe2013introduction,baxter2016exactly,gaudin2014bethe,zheng2024exact}, the transfer matrix formalism, and the Bethe ansatz \cite{bethe1931theorie,faddeev1995algebraic,wierzchucka2024integrability,gritsev2017integrable}.

Existence of QI  or its absence thereof can also be determined  using an alternative approach, particularly when integrals of motion are not known a priori, based on  the Berry-Tabor conjecture or Bohigas-Giannoni-Schmit conjecture \cite{BGS,berry1977level}.  These conjectures provides  a statistical criterion: typical nonintegrable systems follow Wigner-Dyson  statistics after desymmetrization, whereas typical integrable systems follow the Poisson energy level statistics \cite{BGS,berry1977level,d2016quantum}. These conjectures have become a powerful numerical tool for distinguishing between integrable and nonintegrable systems \cite{tekur2020symmetry,rao2020higher}. Furthermore, this approach extends to  Floquet systems \cite{d2014long,ponte2015many}.

Recently the  QI in various systems with long-range and nearest-neighbor interactions has been identified through key signatures like periodicity of the time-evolution unitary operator, and entanglement dynamics, and highly degenerate spectrum or Poisson statistics \cite{yuzbashyan2013quantum,mishra2015protocol,doikou2010introduction,pal2018entangling,naik2019controlled,sharma2024exactly,sharma2024signatures}.  In addition, several other numerical signatures are commonly used to distinguish chaotic from integrable behavior,  such as average adjacent gap ratio, spectral form factor  and the normalized eigenstate entanglement entropy ($ \langle S\rangle/S_{Max}$) \cite{vidmar2017entanglement,hackl2019average,lydzba2020eigenstate,kumari2022eigenstate}.

Motivated by these facts, we investigate the signatures of integrability and chaos in systems with $p$-spin interactions in the infinite-range regime. Specifically, we study the kicked-$p$-spin model, which is a generalized version of Haake’s kicked-top model ($p=2$) \cite{Hakke87,Haakebook}. In Ref.~\cite{munoz2021nonlinear}, the model was generalized in such a way that it becomes extensive and admits a well-defined thermodynamic limit. The two-body case ($p=2$) corresponds to the static Lipkin-Meshkov-Glick (LMG) model in the absence of driving, which exhibits a continuous quantum phase transition between paramagnetic and ferromagnetic phases. More generally, the $p>2$ body case has attracted considerable interest due to its relevance to quantum phase transitions, quantum annealing and optimization, spin-glass, and measurement-based feedback control \cite{Jorg2010,Filippone2011,Bapst2012,Kong2017, Matsuura2017,Munoz-Arias2020}.

In this work, we have analytically calculated the unitary matrix and its time evolution, the eigensystem, the single-qubit reduced density matrix, and entanglement dynamics for arbitrary initial states, any system size $N$, $p$, and $k'$. We find that, for the parameter value $\alpha=\pi$, our $p$-spin model exhibits the signatures of QI for any $k'$, $p$, and $N$. Furthermore, we derive exact expressions for entanglement measures, such as the linear entropy and the entanglement entropy (EE), and find that the model exhibits periodicity when the kicking strength is a rational multiple of $\pi$, i.e., $k' = J\pi$ for rational $J$. The corresponding period depends on $N$ and $p$. However, for the special values of the kicking parameter, periodicity is independent of $N$, and in this regime the model does not admit a well-defined classical limit.

We have also found that the time-evolved unitary operator exhibits periodicity, and its spectrum is highly degenerated, when $k'$ is a  rational multiple of $\pi$. In contrast,  when $k’$ is not a rational multiple of $\pi$, neither the entanglement measures nor the unitary operator exhibits periodicity, and the corresponding eigenvalues are non-degenerate. To conclusively examine the nature of the system for arbitrary  $k'$,$p$, and $N$, we numerically analyze its spectral statistics by computing the higher-order level spacing and higher-order spacing ratio distributions \cite{tekur2020symmetry,rao2020higher}. We observe that the level spacing distribution follows Poisson statistics, indicating the integrable nature of the system. We have numerically computed the average adjacent gap ratio and found that it is consistent with the Poisson value ($2\ln 2 - 1 \approx 0.3863$), suggesting the absence of level repulsion. However, this quantity captures only short-range spectral correlations. To probe long-range correlations in the spectrum, we compute the spectral form factor (SFF) and observe Poisson behavior, indicating the absence of level repulsion.  In addition, we evaluate the  ratio  $ \langle S\rangle/S_{Max}$ and observe that it remains far below $1$ for any $k'$ and $p$. These signatures clearly indicate the integrable nature of our model. We have also observed  similar signatures  of integrability  for the integer multiple of the parameter, i.e. $\alpha=m\pi$ for any $k',N$, $p$ and arbitrary initial state.

The rest of this paper is structured as follows. In Sec. \ref{sec:example-section146}, we present a brief overview of the $p$-spin kicked top model under investigation, highlighting its fundamental aspects.  In Sec. \ref{sec:example-section46}, we present an exact analytical solution for entanglement measures, such as linear entropy and EE for the parameter values $\alpha=\pi$,  any $k'$, $N$ and for arbitrary initial unentangled states for any even-$p$. In Sec. \ref{sec:example-section49}, we derive similar expressions for any odd-$p$.  In Sec. \ref{sec:example-section48}, we provide extensive numerical results by analyzing the spectral statistics, average adjacent gap ratio and spectral form factror for any $k'$ and $p$. In Sec. \ref{sec:example-section4888}, we numerically calculated the average eigenstate EE  for half-bipartition and its behavior with parameters $k',N, p$, and $\alpha=m\pi$. In Sec. \ref{sec:example-section6}, we provide a summary of the main results and conclusions of our work.

\section{ The $p$-QKT model}\label{sec:example-section146}
The Hamiltonian governing the dynamics of the generalized kicked-top model with
$p$-spin interactions is given by \cite{munoz2021nonlinear}
\begin{equation}
\label{Eq:QKT4}
H(t)= \frac{\alpha}{\tau}\hat{J}_y+\frac{k'}{p J^{p-1}} \hat{J}^{p}_z\sum_{n = -\infty}^{ \infty} \delta(t-n\tau),
\end{equation}
The parameter \( \tau \) denotes the time interval between successive periodic kicks, and \( {J}_{x,y,z} \) are the components of the angular momentum operator \( \mathbf{J} \). Here, the parameter $\alpha$ represents a rotation around the $y$-axis.  The parameter $p$ determines the order of interaction and governs the degree of nonlinearity, with larger $p$ corresponding to higher-order multi-spin interactions. The parameter  $k'$ quantifies the strength of a twist introduced between successive kicks and it controls chaos. In the absence of this twist ($k'=0$), the dynamics reduce to a simple rotation, which is integrable. The Floquet operator corresponding to the Hamiltonian in Eq.~(\ref{Eq:QKT4}) is given by
\begin{eqnarray} \label{Eq:QKT1}
\mathcal{U} &=& \exp\left (- \frac{i~k'}{p j^{p-1}} {J}^{p}_z \right) ~ \exp\left(-i \alpha {J}_y\right).
\end{eqnarray}
 A connection to a many-body system can be established  by considering $\bf{J}$ spin as the total of $N=2j$ spin-$1/2$ qubits, replacing ${J}_{(x,y,z)}$  with $\sum_{l=1}^{N}\sigma_l^{x,y,z}/2$. Here, $\sigma_l^{x,y,z}$ are the standard Pauli matrices \cite{Hakke87,Arecchi1972,ExactDogra2019,sharma2024exactly,kumari2022eigenstate}. The corresponding Floquet operator  can   then be expressed as,
\begin{eqnarray} \label{Eq:QKT2}
\mathcal{U} &=& \exp\left (- \frac{i~k'}{p j^{p-1}} \left( \sum^N_{ l=1}\frac {\sigma_l^{z}}{2} \right)^p  \right)  \exp\left(\frac{-i~ \alpha}{2} \sum_{l=1}^{N}\sigma_l^y \right).
\end{eqnarray}
The
$p$th power of the collective spin operator can be expanded in terms of products of Pauli operators as follows;
\begin{eqnarray}
 \left(\sum^N_{ l=1}\frac {\sigma_l^{z}}{2} \right)^p=\frac{1}{2^p}\sum^{\mbox{min}(r,p)}_{r=0} a_r(N,p) \sum_{1\leq i_1<...i_r<N} \sigma^z_{i_1} .....\sigma^z_{i_r},
\end{eqnarray}
where, the coefficient $a_r(N,p)$ can be calculated as follows:
\begin{eqnarray}\nonumber
 a_r(N,p)=\frac{1}{2^N}\sum^r_{v=0} \sum^{N-r}_{w=0} (-1)^{(r-j)}\binom{r}{v} \binom{N-r}{w} (2v+2w-N)^p.
\end{eqnarray}
Here, $r$ denotes the order of the multi-spin interaction. The coefficient $a_r(N,p)$ is nonzero only when $r$ and $p$ have same parity. Furthermore, $a_r(N,p)=0$ for $r>p$.

The classical limit is obtained by taking  $N\rightarrow \infty$ for fixed $k'$, which yields a well-defined classical map. The dynamics is described by a map on the surface of the unit sphere in phase space, defined by \( X^2 + Y^2 + Z^2 = 1 \), where the variables are given by \( X = {J}_x/j \), \( Y = {J}_y/j \), and \( Z = {J}_z/j \).
 The classical map of the $p$-spin kicked top for $\alpha=\pi$, is given as follows \cite{munoz2021nonlinear}:
\begin{align}
X_{m+1} &=-X_m \cos\!\left(k'(-Z_m)^{p-1}\right) -  Y_m\sin\!\left(k'(-Z_m)^{p-1}\right), \nonumber \\
Y_{m+1} &= -X_m\sin\!\left(k'(-Z_m)^{p-1}\right) +  Y_m\cos\!\left(k'(-Z_m)^{p-1}\right) , \nonumber \\
Z_{m+1} &= -Z_m.
 \end{align}
\label{eq:ClassicalMap}
\begin{figure}[t]
\includegraphics[width=0.47\textwidth]{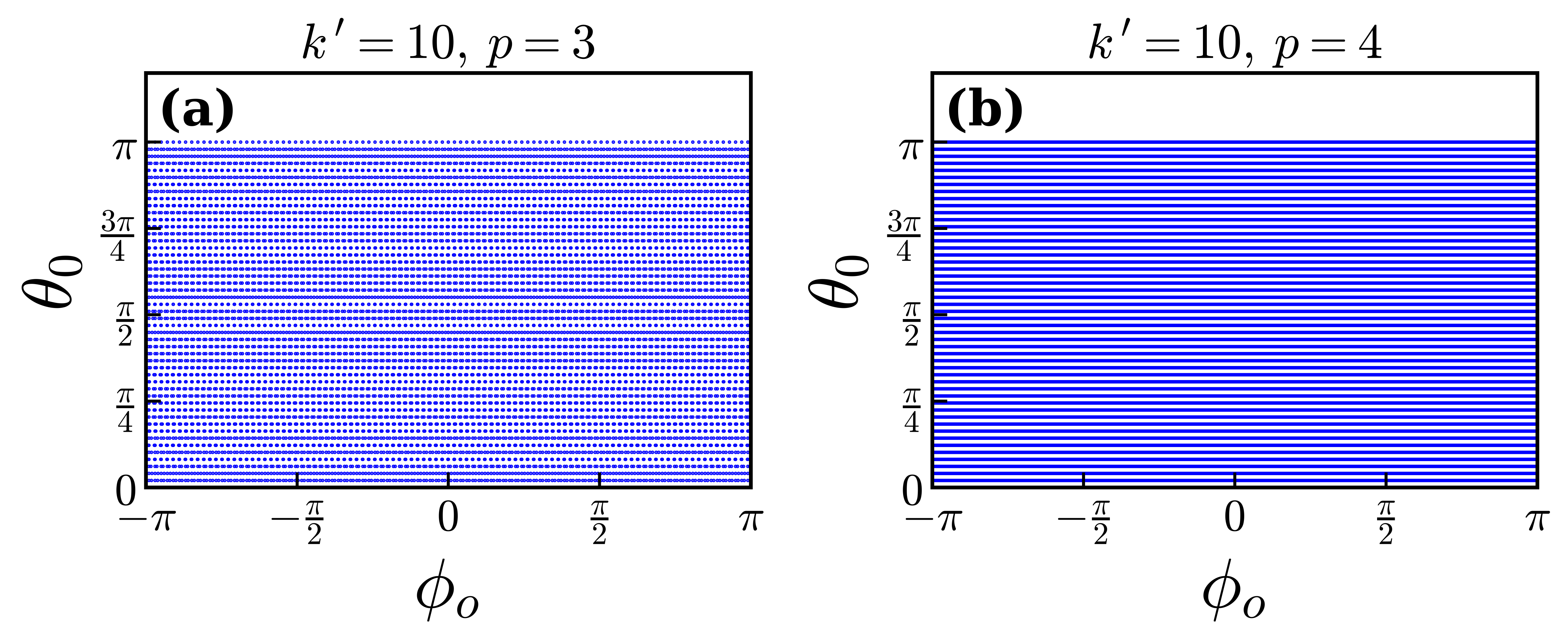}
\caption{Phase-space pictures of the classical $p$-spin kicked top for $\alpha=\pi$.}
\label{fig:degenrated211}
\end{figure}
In Ref. \cite{UdaysinhPeriodicity2018}, the classical map for various values of $\alpha$ and $p=2$, has been studied extensively. In the QKT model, a transition from a regular region to a chaotic sea has been observed for different values of $\alpha$ with $k'$ \cite{Hakke87,UdaysinhPeriodicity2018,munoz2021nonlinear}. In various studies, the LLE is used to examine the chaotic behavior of the system and to observe the transition from a regular to  chaos \cite{wang2021multifractality,munoz2021nonlinear}. This is also reflected in the corresponding phase-space \cite{UdaysinhPeriodicity2018}. It has been shown analytically that the LLE is  zero for $\alpha=m\pi$, where $m$ is an integer \cite{UdaysinhPeriodicity2018,munoz2021nonlinear}.  This is reflected in the phase space shown in Fig. \ref{fig:degenrated211} as regular dynamics, which implies that the kicked top model is  classically integrable for any $k', p$ and $\alpha=m\pi$ \cite{wang2021multifractality,UdaysinhPeriodicity2018,munoz2021nonlinear}. For other values of $\alpha\neq m\pi$, the LLE  for  $p$-body interaction can be calculated using $\ln\!\big[k'(p-1)\sin(\alpha)\big] - (p-1)$, which shows the transition from regular to chaotic sea with $k'$ \cite{wang2021multifractality,munoz2021nonlinear}. For a fixed $k'$ and $p$, the system exhibits maximum chaotic behavior at $\alpha = \pi/2$ \cite{Hakke87}.  However, if the parameter $k'\propto N^{p-1}$, then in the limit $N\rightarrow \infty$, the classical limit of the model does not exist for any $\alpha$.

An important question is whether quantum integrability (QI) persists in the PQKT model for parameters $k'$, $N$, $p$, and $\alpha = \pi$.
We address this question  analytically and numerically using various signatures of QI.
These include periodic behavior in entanglement and operator dynamics, spectral statistics, the spectral form factor (SFF), and normalized eigenstate entanglement entropy (EEE).
Our results shows that QI exists for these parameters.
To obtain an analytical solution for the system with any  $N$ qubits, we use the general basis presented in Ref. \cite{sharma2024exactly}.
 When $N$ is even, the basis is given as follows:
\begin{eqnarray}\label{Eq:evenBasis1}
\begin{split}
\ket{\phi_q^{\pm}}&=\frac{1}{\sqrt{2}}\left(\ket{w_q}\pm {(-1)^{\left(j-q\right)}} \ket{\overline{w_q}}\right), 0\leq q\leq j-1\\
\mbox{and} &\;\;
\ket{\phi_{\frac{N}{2}}^+}=\left({1}/{\sqrt{\binom{N}{\frac{N}{2}}}}\right)\sum_{\mathcal{P}}\left(\otimes^{\frac{N}{2}}\ket{0}\otimes^{\frac{N}{2}}\ket{1}\right)_\mathcal{P},
\end{split}
\end{eqnarray}
whereas the basis for odd $N$  is given as,
\begin{equation}\label{Eq:oddBasis1}
\ket{\phi_q^{\pm}}=\frac{1}{\sqrt{2}}\left(\ket{w_q}\pm {i^{\left(N-2q\right)}} \ket{\overline{w_q}}\right),
0\leq q\leq \dfrac{N-1}{2};
\end{equation}
where
$\ket{w_q}=\left({1}/{\sqrt{\binom{N}{q}}}\right)\sum_\mathcal{P}\left(\otimes^q \ket{1} \otimes^{(N-q)}\ket{0}\right)_\mathcal{P}$ and
$\ket{\overline{w_q}}=\left({1}/{\sqrt{\binom{N}{q}}}\right)\sum_\mathcal{P}\left(\otimes^{q}\ket{0}\otimes^{(N-q)}\ket{1}\right)_\mathcal{P}$.
Both states are definite particle-number states. The $\sum_\mathcal{P}$ denotes the sum over all possible permutations.
These basis states $\ket{\phi_j^{\pm}}$ are  the eigenstate of the  parity operator having eigenvalues $\pm1$ i.e. $\otimes_{l=1}^{N}\sigma_l^y\ket{\phi_j^{\pm}}=\pm\ket{\phi_j^{\pm}}$. The coherent state provides a convenient initial state that connects quantum and classical dynamics. As a minimum-uncertainty state localized on the Bloch sphere, it is suitable for exploring the semiclassical behavior of the system. The coherent state \cite{puri2001mathematical} is given in the computational basis as follows:
\begin{equation}\label{Eq:oddBasis11}
\ket{\psi_0}=|\theta_0,\phi_0\rangle
=\cos(\theta_0/2)|0\rangle+e^{-i\phi_0}\sin(\theta_0/2)|1\rangle.
\end{equation}
Although the computational basis has dimension $2^N$, permutation symmetry allows the system to be expressed in the generalized basis introduced above, reducing the Hilbert-space dimension to $N+1$. Thus, only $N+1$ coefficients are required to describe the state and its time evolution. Moreover, a significant advantage of using $\ket{\phi}$ basis is that the unitary operator becomes block-diagonal.  This advantage will be discussed in details in the subsequent part of this paper. Throughout this paper, we consistently use the parameter value  $\alpha = \pi$ unless otherwise stated.
\section{For even $p$-body interaction}\label{sec:example-section46}
In this section, we have analytically calculated the unitary operator, its time evolution, reduced density matrix (RDM) of single qubit, and the entanglement dynamics for arbitrary initial states with any even-$N$ and even-$p$-spin. To quantify the entanglement dynamics, we utilize linear entropy and EE.  The floquet opertor for even values of  $p$ is invarient under $R_y(\pi)$, therefore, it can be written in a block-diagonal structure \cite{munoz2021nonlinear}. In the $\ket{\phi}$ basis, the unitary operator $\mathcal{U}$ for even-$p$ decomposes into two blocks, $\mathcal{U}_+$ and $\mathcal{U}_-$, corresponding to the even and odd-parity sectors, respectively.. An advantage of the parameter value $\alpha = \pi$ is that these two blocks become diagonal matrices, with all off-diagonal elements equal to zero.
\subsection{For even qubits}
The two blocks for even-$N$, $\mathcal{U}_+$ and $\mathcal{U}_-$ having dimensions $j+1$ and $j$, in the $\ket{\phi}$ basis  can be written as follows:
\begin{eqnarray}
 \mathcal{U}_+&=&\text{diag}[(i)^N~ f_q]; ~~~0\leq q\leq N/2, ~~~\mbox{and}\\
 \mathcal{U}_-&=&\text{diag}[(i)^{N-2}~ f_q]; ~~~0\leq q\leq \dfrac{N-2}{2},
\end{eqnarray}
where $f_q$ for even-$N$ qubits can be expressed as follows:
\begin{equation}\label{Eq:per21}
 f_q=\exp\left[\frac{-i k'}{p~j^{p-1}}\left(\frac{(N-2q)^p-a_{0}(N,p)}{2^p}\right)\right].
\end{equation}
 Here, the coefficient $a_{0}(N,p)$ can be calculated as follows:
\begin{eqnarray}\label{Eq:per22}
 a_{0}(N,p)=\frac{1}{2^N} \sum^{N}_{w=0}  \binom{N}{w} (2w-N)^p.
\end{eqnarray}
The eigenvalues  of $\mathcal{U}_+$ ~($\mathcal{U}_-$) are given by  $(i)^N~ f_q$ ~($(i)^{N-2}~ f_q$), where $q$ varies in the interval $[0,N/2]$ $\left([0,(N-2)/2]\right)$. The eigenvectors corresponding to these eigenvalues are $v_l=[0,0,\dots,1\dots,0]^T $, where $v_l$ represents the  basis vector with $1$ in the $l$th position and $l \in \{0,1,\dots,N\}$. The diagonal structure of the matrices $\mathcal{U}_+$ and $\mathcal{U}_-$ simplifies the analytical computations for any $N$ and $k'$. This structure  further simplifies the calculation of the time evolution. The $n$th time evolution of two blocks $\mathcal{U}_+$ and $\mathcal{U}_-$ can be written as follows:
\begin{eqnarray}
 \mathcal{U}^{n}_+&=&\text{diag}(A_q^+); ~~~0\leq q\leq N/2,~~~\mbox{and}\\
 \mathcal{U}^{n}_-&=&\text{diag}(A_q^-); ~~~0\leq q\leq \dfrac{N-2}{2},
\end{eqnarray}
where $A_q^+$, $A_q^-$  and $f^n_q $ can be expressed as follows:
\begin{eqnarray}
 A^+_q&=&((i)^N)^n~f^n_q  ; ~~~~~~0\leq q\leq N/2 \label{Eq:per2} \\
A^-_q&=&((i)^{N-2})^n~f^n_q ; ~~~~~~0\leq q\leq \dfrac{N-2}{2}, \label{Eq:per1}\\
f^n_q&=&\exp\left[\frac{-i~ n~k'}{p~j^{p-1}}\left(\frac{(N-2q)^p-a_{0}(N,p)}{2^p}\right)\right].\label{Eq:per23}
\end{eqnarray}
 We initialize each of the $N$ qubits in the coherent state $\ket{\psi_0}$, resulting in an unentangled product state for the $N$-qubit system, given by $\ket{\psi} = \ket{\psi_0}^{\otimes N}$.
As mentioned earlier, the dimensionality of the system can be reduced by an  basis transformation. After the basis transformation, the arbitrary initial state $\ket{\psi}$ for any even number of qubits in $\ket{\phi}$ basis can be expressed as:
\begin{equation}
 \ket{\psi}= \sum_{q=0}^{N/2-1}\frac{1}{\sqrt{2}}\left( a_{q+1} \ket{\phi_{q}^+} +b_{q+1} \ket{\phi_{q}^-}\right)+a_{\frac{N+2}{2}} \ket{\phi_{\frac{N}{2}}^+},
\end{equation}
where the coefficients $a_{q+1}$, $b_{q+1}$ and $a_{\frac{N+2}{2}}$ are given as follows \cite{sharma2025exact}:
\begin{eqnarray}\nonumber \label{Eq:arbitaray11}
a_{q+1}&=&\sqrt{\binom{N}{q}}\left(\cos^{N-q}\left(\theta_0/2\right) e^{-i q\phi_0} \sin^{q}\left(\theta_0/2\right) +~i^{N-2q}\right. \\  &&\left.\cos^{q}\left(\theta_0/2\right) e^{-i (N-q)\phi_0}\sin^{N-q}\left(\theta_0/2\right)\right),\\ \label{Eq:arbitaray12} \nonumber
b_{q+1}&=&\sqrt{\binom{N}{q}}\left(\cos^{N-q}\left(\theta_0/2\right) e^{-i q \phi_0} \sin^{q}\left(\theta_0/2\right) -~i^{N-2q}\right. \\   &&\left.\cos^{q}\left(\theta_0/2\right) e^{-i (N-q)\phi_0}\sin^{N-q}\left(\theta_0/2\right)\right)~~\mbox{and}\\  \label{Eq:arbitaray14}
a_{\frac{N+2}{2}}&=&\sqrt{\binom{N}{\frac{N}{2}}}\left(e^{-i\frac{N}{2} \phi_0}\cos^{\frac{N}{2}}\left(\theta_0/2\right)  \sin^{\frac{N}{2}}\left(\theta_0/2\right)\right).
\end{eqnarray}
Here, $q$ lies in the interval $[0, \frac{N-2}{2}]$.
The state $\ket{\psi_n}$ is obtained after the $n$th application of the unitary operator $\mathcal{U}$ on the initial state $\ket{\psi}$, and can be expressed as follows:
\begin{eqnarray}
 \ket{\psi_n}&=&\mathcal{U}^n \ket{\psi}\\ \nonumber
 &=&\sum_{q=0}^{N/2-1}\left(B^+_q\ket{\phi_{q}^+}+B^-_q\ket{\phi_{q}^-}\right)+B^+_\frac{N}{2}\ket{\phi_{\frac{N}{2}}^+},
\end{eqnarray}
\begin{figure}[t]\vspace{0.4cm}
\includegraphics[width=0.47\textwidth]{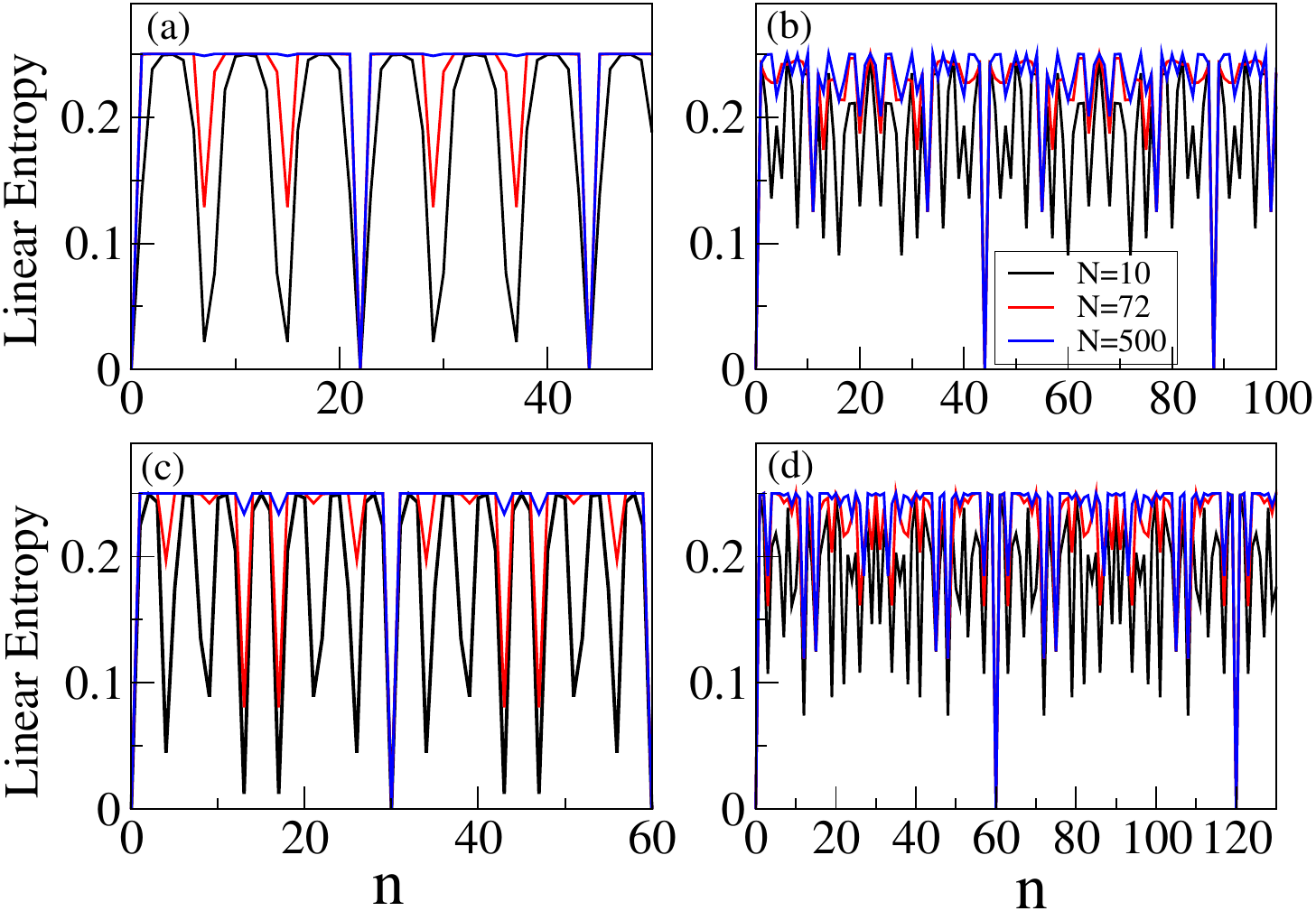}
\caption{The linear entropy is plotted using Eq.~(\ref{EQ:Entropy128}) for even-$N$ and  the  initial states $\ket{\pi/4,-\pi/4}$. Here, $p=2$ ($p=4$) in (a) and (c) ((b) and (d)). Whereas $J=3/11$ ($J= 7/15$)   in (a) and (b) ((c) and (d)) and  $k'=J j^{p-1} \pi$. }
\label{fig:8qubitavg}
\end{figure}
where the coefficients $B^+_q$, $B^-_q$ and $B^+_{{N}/{2}}$ can be written as:
\begin{eqnarray}\nonumber \label{Eq:11}
 B^+_q&=&A^+_q ~a_{q+1}{/}\sqrt{2},\quad
  B^-_q=A^-_q ~b_{q+1}{/}\sqrt{2}\quad \mbox{and}\\
   B^+_{{N}/{2}}&=&A^+_{{N}/{2}} ~a_{\frac{N+2}{2}}.\
\end{eqnarray}
The single-qubit RDM is then given as follows:
\begin{equation}
\rho_1(n)=\frac{1}{2}\left(
\begin{array}{cc}
 r_n & \bar{w}_n \\
\bar{w}_n^* & 2-r_n \\
\end{array}
\right),
\end{equation}
where the coefficient $r_n$ and $\bar{w}_n$ are given as follows:
\begin{eqnarray} \nonumber
 r_n&=&{1}+\sum_{q=1}^{N/2-1}\left\lbrace\left[{\binom{N-1}{q}-\binom{N-1}{q-1}}\right]\Big{/}{\binom{N}{q}}\right\rbrace  \left[B^+_{q} (B^-_{q})^* \right. \\  && \left. + B^-_{q} (B^+_{q})^*\right]+B^+_{0} (B^-_{0})^*+B^-_{0} (B^+_{0})^* ~~ \mbox{and} \label{Eq:arbitaray125} \\  \nonumber
 \vspace{1cm}
 \bar{w}_n&=&\sum_{q=1}^{N/2-1}\left\lbrace{\binom{N-1}{q}}\Big{/}\left[\sqrt{{\binom{N}{q}}{\binom{N}{q+1}}}\right]\right\rbrace\left[\left(B^+_{q}+ B^-_{q}\right)\right.\\ \nonumber &&\left.\left((B^+_{q+1})^* +(B^-_{q+1})^*\right)-\left(B^-_{q+1}- B^+_{q+1}\right)\left((B^-_{q})^* \right.\right.\\ \nonumber &&\left.\left.-(B^+_{q})^*\right)\right]+\left\lbrace\sqrt{2}~{\binom{N-1}{\frac{N-2}{2}}}\Big{/}\left[\sqrt{{\binom{N}{\frac{N-2}{2}}}{\binom{N}{N/2}}}\right]\right\rbrace \\ \nonumber  &&\left[\left(B^+_{\frac{N-2}{2}}+ B^-_{\frac{N-2}{2}}\right)\left(B^+_{{N}/{2}}\right)^*+\left(\left(B^-_{\frac{N-2}{2}}\right)^*\right.\right.\\  &&\left.\left.-\left( B^-_{\frac{N-2}{2}}\right)^*\right)B^+_{{N}/{2}}\right].\label{Eq:arbitaray126}
\end{eqnarray}
The eigenvalues $\lambda_1$ and $\lambda_2$ of single qubit, $\rho_1(n)$ are $\frac{1}{2} \left(1\pm\sqrt{1-{r}_n\left(2-{r}_n\right)-|\bar{{w}}_n}|^2\right)$. The linear entropy of single-qubit RDM is given as follows:
\begin{equation}\label{EQ:Entropy128}
 S_{(\theta_0,\phi_0)}^{(N)}(n,k',p)=[r_n(2-r_n)-|{\bar{w}_n}|^2]/2.
\end{equation}
We compute the linear entropy and analyze its time dependence for arbitrary initial states and any value of $k'$ and $n$ using Eq.~(\ref{EQ:Entropy128}). The entanglement entropy (EE) is given by $-\left[\lambda_1\ln(\lambda_1)+\lambda_2\ln(\lambda_2)\right]$, where $\lambda_1$ and $\lambda_2$ are the eigenvalues of the RDM. To systematically analyze the entanglement dynamics, we consider the following cases:

\textit{\textbf{Case I}: $k'\propto j^{p-1}$.}
In this case  the classical limit of the system does not exist and system is purely quantum. There are further two subcases (a) and (b).

\textit{(a)} The case  ${k'}={J \pi ~j^{p-1}}$, where $J=r/h$  such that GCD $(r,h)=1$: we observe periodic entanglement dynamics for any even-$N$ and even-$p$, with period $ph$. This is shown in Fig.~\ref{fig:8qubitavg}.

\begin{figure}[t]\vspace{0.4cm}
\includegraphics[width=0.47\textwidth]{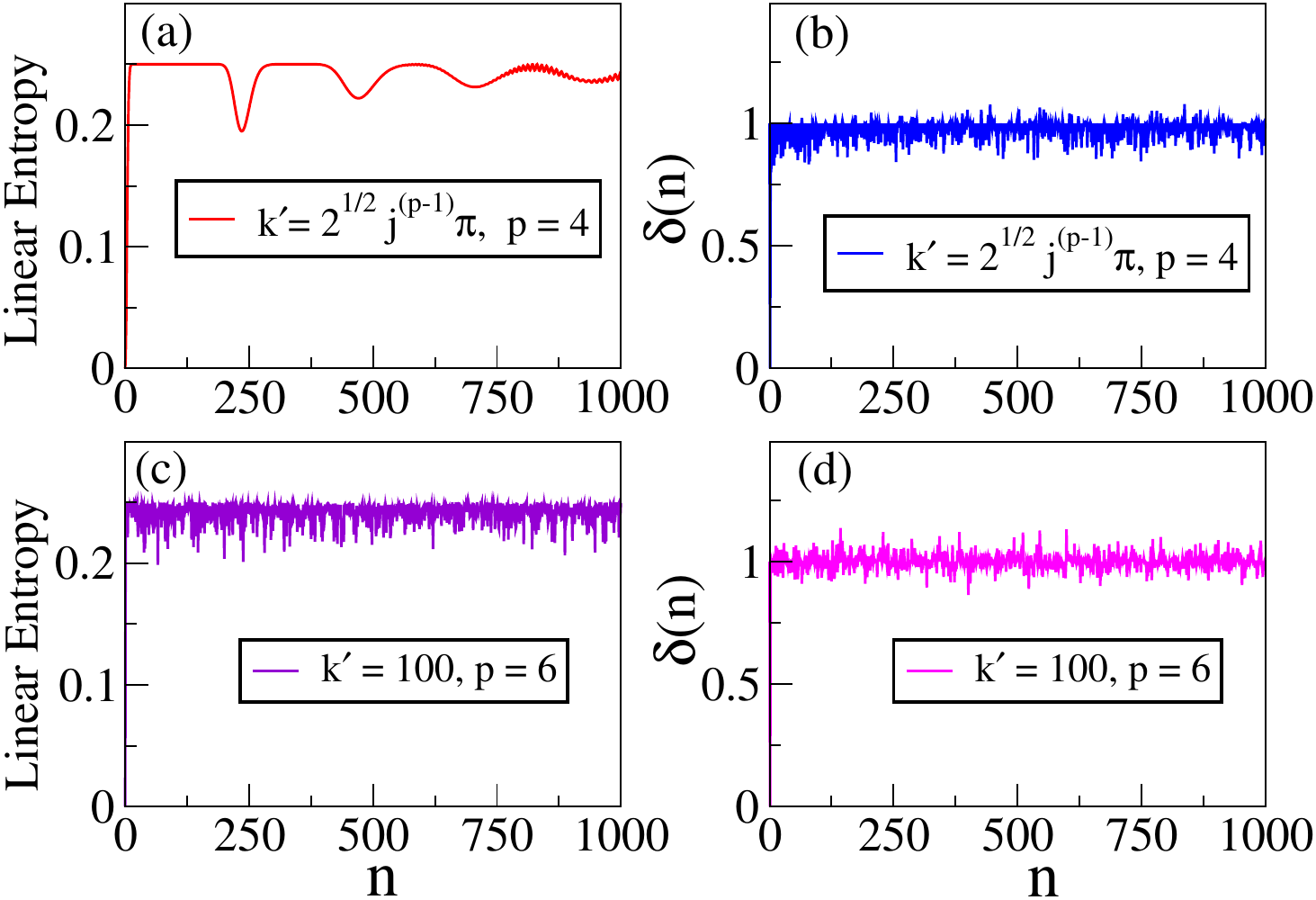}
\caption{The linear entropy and deviation $\delta(n)$ are shown  for $k'=\sqrt{2} j^{p-1}\pi$ and $p=4$ (upper panel), and for $k'=100$, $p=6$ (lower panel) and $N=500$.}
\label{fig:spectrum35}
\end{figure}

\textit{(b) $k'$ is not a rational multiple of $\pi$:} In this scenario  the periodic behavior of entanglement is absent. The results are shown in Fig.~\ref{fig:spectrum35}~a.

\textit{\textbf{Case II}: $k'$ is finite and independent of $j$:} In this case, the classical limit of the system exists. There are further two subcases (a) and (b).

\textit{(a) $k'$ is a rational multiple of $\pi$:}

 When $k'=J\pi$, we observe periodic entanglement dynamics. In contrast to Case I~(a), the period depends on both $N$ and $p$ (results are not shown here).

\textit{(b) $k'$ is not a rational multiple of $\pi$:} Here  the periodic behavior is absent. Results are shown in Fig.~\ref{fig:spectrum35}~c.

Using this approach, we additionally present detailed analytical calculations and entanglement measures for a special class of initial states, $\ket{0, \phi_0}$ and $\ket{\pi/2, \pm \pi/2}$, in \ref{appendix:A}.
Additionally, we find that for certain initial states, such as $\ket{0, \phi_0}$ and $\ket{\pi, \pm \pi}$, no entanglement is generated between the subsystems (See~\ref{appendix:A} for a detailed calculation).

\subsection{For odd qubits}
\begin{figure}[t]\vspace{0.1cm}
\includegraphics[width=0.47\textwidth] {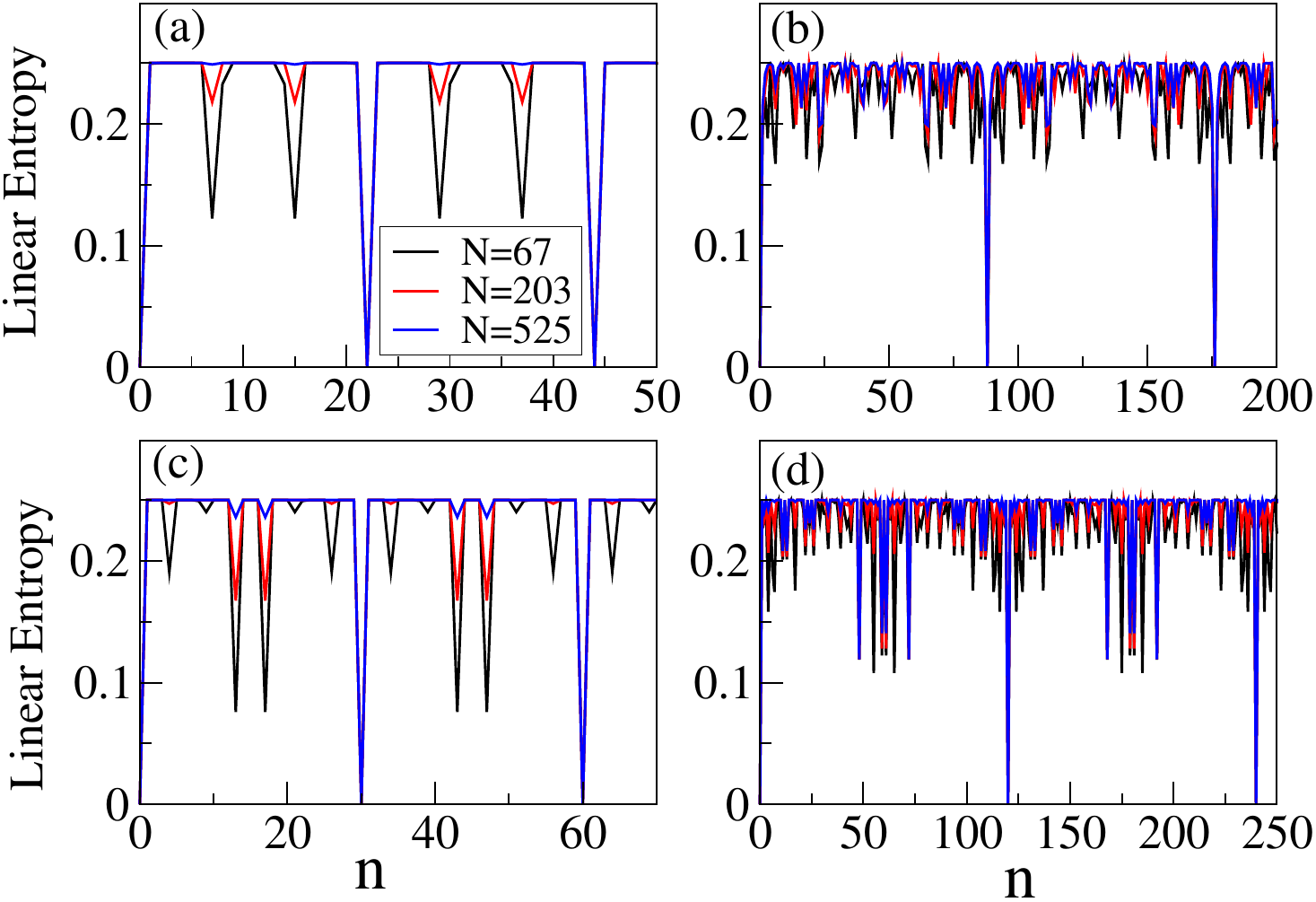}
\caption{ Same as Fig. \ref{fig:8qubitavg} but for odd-$N$.}
\label{fig:8qubitavg1}
\end{figure}

Now, we perform a similar analysis for odd $N$ and even-$p$ as in Sec. \ref{sec:example-section46}~A and present the corresponding results. The unitary operator $\mathcal{U}$  can be expressed in two blocks $\mathcal{U}_+$ and $\mathcal{U}_-$ as follows:
\begin{eqnarray}
 \mathcal{U}_+&=&\text{diag}[(i)^N~ f_q]; ~~~0\leq q\leq \dfrac{N-1}{2}~~\mbox{and}\\
 \mathcal{U}_-&=&\text{diag}[(i)^{N-2}~f_q]; ~~~0\leq q\leq \dfrac{N-1}{2},
\end{eqnarray}
where $f_q, a_0(N,p)$ and $f^n_q$ are defined in Eqs. (\ref{Eq:per21}),  (\ref{Eq:per22}) and  (\ref{Eq:per23}) respectively. The $n$th time evolution of two blocks $\mathcal{U}_+$ and $\mathcal{U}_-$ can be written as,
\begin{eqnarray}
 \mathcal{U}^{n}_+&=&\text{diag}(A_q^+); ~~~0\leq q\leq \dfrac{N-1}{2}~~\mbox{and}\\
 \mathcal{U}^{n}_-&=&\text{diag}(A_q^-); ~~~0\leq q\leq \dfrac{N-1}{2},
\end{eqnarray}
where $A_q^+$ and $A_q^-$ can be expressed as:
\begin{eqnarray}
 A^+_q&=&((i)^N)^n~f^n_q ; ~~~0\leq q\leq \dfrac{N-1}{2} \quad \mbox{and}\label{EQ:Entropy26}\\
A^-_q&=&((i)^{N-2})^n~f^n_q ; ~~~0\leq q\leq \dfrac{N-1}{2}. \label{EQ:Entropy56}
\end{eqnarray}
 In this case, the arbitrary initial state  can be expressed as follows:
\begin{equation}
 \ket{\psi}= \sum_{q=0}^{\frac{N-1}{2}}\frac{1}{\sqrt{2}}\left( \tilde{a}_{q+1} \ket{\phi_{q}^+} +\tilde{b}_{q+1} \ket{\phi_{q}^-}\right),
\end{equation}
where the coefficients $\tilde{a}_{q+1}$ and  $\tilde{b}_{q+1}$ are given as follows, with $q \in [0,\frac{N-1}{2}]$:
\begin{eqnarray}\nonumber \label{Eq:arbitaray}
\tilde{a}_{q+1}&=&\sqrt{\binom{N}{q}}\left(\cos^{N-q}\left(\theta_0/2\right) e^{-i q\phi_0} \sin^{q}\left(\theta_0/2\right) -~i^{N-2q}\right. \\  &&\left.\cos^{q}\left(\theta_0/2\right) e^{-i (N-q)\phi_0}\sin^{N-q}\left(\theta_0/2\right)\right)~~\mbox{and}\\ \label{Eq:arbitaray1} \nonumber
\tilde{b}_{q+1}&=&\sqrt{\binom{N}{q}}\left(\cos^{N-q}\left(\theta_0/2\right) e^{-i q \phi_0} \sin^{q}\left(\theta_0/2\right) +~i^{N-2q}\right. \\   &&\left.\cos^{q}\left(\theta_0/2\right) e^{-i (N-q)\phi_0}\sin^{N-q}\left(\theta_0/2\right)\right).\\ \nonumber
\end{eqnarray}
The state $\ket{\psi_n}$ can then be obtained by the $n$ successive application of the unitary operator $\mathcal{U}$ on the arbitrary initial state $\ket{\psi}$, expressed as,
\begin{eqnarray}\nonumber
 \ket{\psi_n}=\mathcal{U}^n \ket{\psi}
 =\sum_{q=0}^{N/2-1}\left(B^+_q\ket{\phi_{q}^+}+B^-_q\ket{\phi_{q}^-}\right),
\end{eqnarray}
where the coefficients $B^+_q$ and  $B^-_q$ can be written as,
\begin{eqnarray} \label{Eq:12}
 B^+_q&=&A^+_q ~\tilde{a}_{q+1}{/}\sqrt{2}~~\mbox{and}~~B^-_q=A^-_q ~\tilde{b}_{q+1}{/}\sqrt{2}.\\ \nonumber
\end{eqnarray}
The single-qubit RDM is then given as follows:
\begin{equation}
\rho_1(n)=\frac{1}{2}\left(
\begin{array}{cc}
 t_n & \bar{m}_n \\
\bar{m}_n^* & 2-t_n \\
\end{array}
\right),
\end{equation}
where the coefficients $t_n$ and $\bar{m}_n$ are given as follows:
\begin{eqnarray}\nonumber
 t_n&=&{1}+\sum_{q=1}^{\frac{N-1}{2}}\left\lbrace\left[{\binom{N-1}{q}-\binom{N-1}{q-1}}\right]\Big{/}{\binom{N}{q}}\right\rbrace \left[B^+_{q} (B^-_{q})^*\right.\\  &&\left.+B^-_{q} (B^+_{q})^*\right]+B^+_{0} (B^-_{0})^*+B^-_{0} (B^+_{0})^*~~~~ \label{Eq:arbitaray129}\mbox{and}  \\ [0.25cm]  \nonumber
 \bar{m}_n&=&\sum_{q=0}^{\frac{N-3}{2}}\left\lbrace{\binom{N-1}{q}}\Big{/}\left[\sqrt{{\binom{N}{q}}{\binom{N}{q+1}}}\right]\right\rbrace\left[\left(B^+_{q}+ B^-_{q}\right)\right.\\ \nonumber &&\left.\left((B^+_{q+1})^* +(B^-_{q+1})^*\right)-\left(B^-_{q+1}- B^+_{q+1}\right)\left((B^-_{q})^* \right.\right.\\ \nonumber &&\left.\left.-(B^+_{q})^*\right)\right] -~\left\lbrace i~{\binom{N-1}{\frac{N-1}{2}}}\Big{/}\left[\sqrt{{\binom{N}{\frac{N-1}{2}}}{\binom{N}{\frac{N+1}{2}}}}\right]\right\rbrace \\ &&\left[\left(B^+_{\frac{N+1}{2}}+ B^-_{\frac{N+1}{2}}\right)\left(\left(B^+_{\frac{N+1}{2}}\right)^*-\left( B^-_{\frac{N+1}{2}}\right)^*\right)\right].
 \label{Eq:arbitaray128}
\end{eqnarray}
The eigenvalues of $\rho_1(n)$ are $\frac{1}{2} \left(1\pm\sqrt{1-{t}_n\left(2-{t}_n\right)-|\bar{{m}}_n}|^2\right)$. The linear entropy of single qubit is given as follows:
\begin{equation}\label{Eq:entropyy6}
 S_{(\theta_0,\phi_0)}^{(N)}(n,k',p)=[t_n(2-t_n)-|{\bar{m}_n}|^2]/2.
\end{equation}
The linear entropy can be calculated using Eq.~(\ref{Eq:entropyy6}) for any $k'$ and $n$, and for arbitrary initial states with odd $N$ and even $p$. As discussed above, in Case~I~(a), the entanglement measures exhibit periodicity for even $N$. Extending this analysis to odd $N$, we find that the linear entropy is periodic, with a period of $2^{p-1}h$. This is shown in Fig.~\ref{fig:8qubitavg1}. Periodic entanglement dynamics are also observed in Case~II~(a), with the period depending on $N$, $p$, and $h$.  In contrast, for Cases~I~(b) and II~(b), the periodic behavior of entanglement is absent for odd-$N$ as well.

Using the same procedure, we  additionally provide detailed calculations for a special class of initial states in \ref{appendix:B}. We find that certain initial states, such as $\ket{0,\phi_0}$ and $\ket{\pi,\pm\pi}$, do not generate entanglement between the subsystems for any $n$. A similar behavior was observed in the previous section for even-$N$.

\begin{figure}[t]\vspace{0.1cm}
\includegraphics[width=0.47\textwidth]{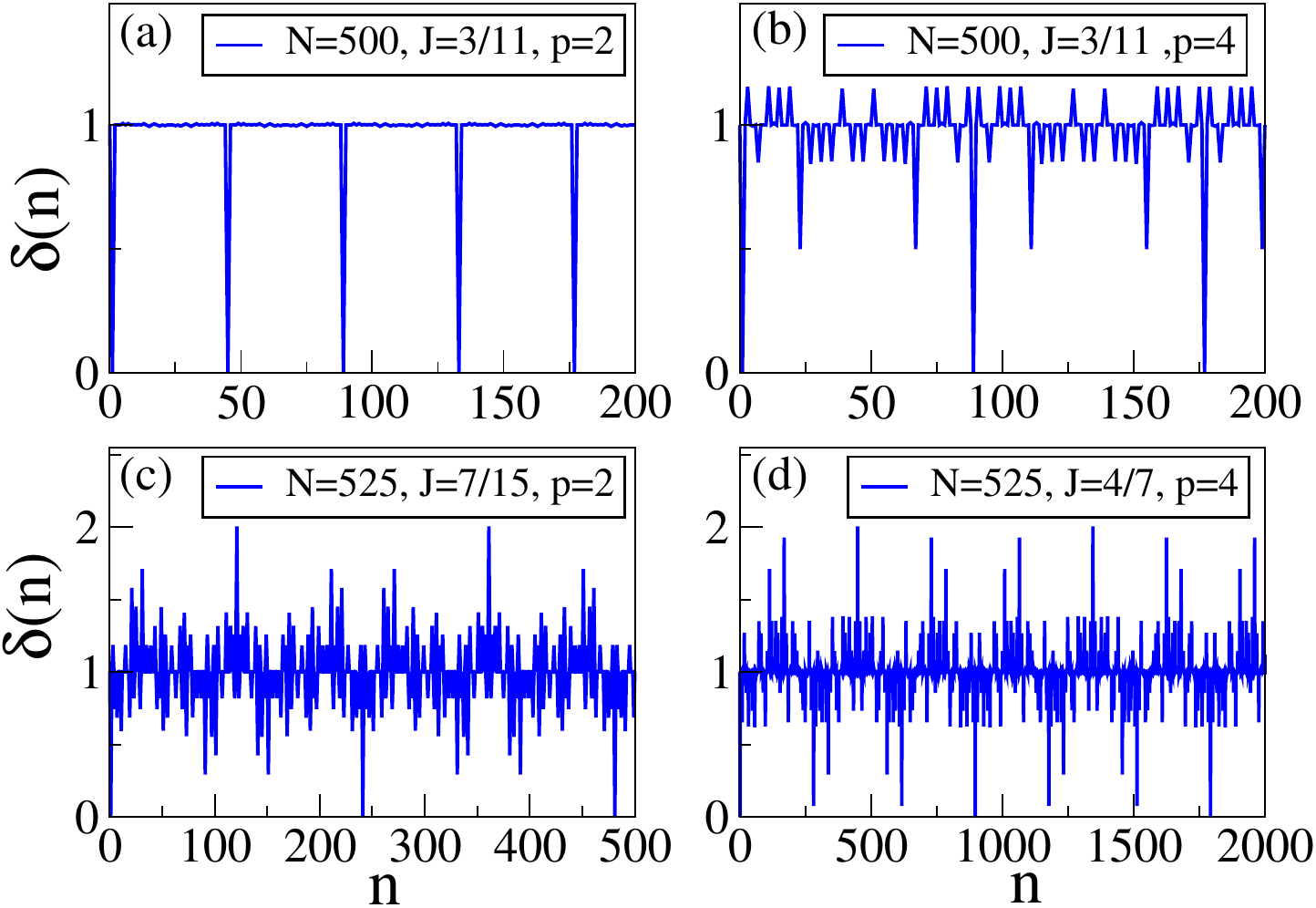}
\caption{Deviation $\delta(n)$ for the rational values of $J$ and $k'=J j^{p-1} \pi$ with $N$ and $p$.}
\label{fig:degenrated2111}
\end{figure}

\begin{figure}[htbp]\vspace{0.3cm}
\includegraphics[width=0.47\textwidth]{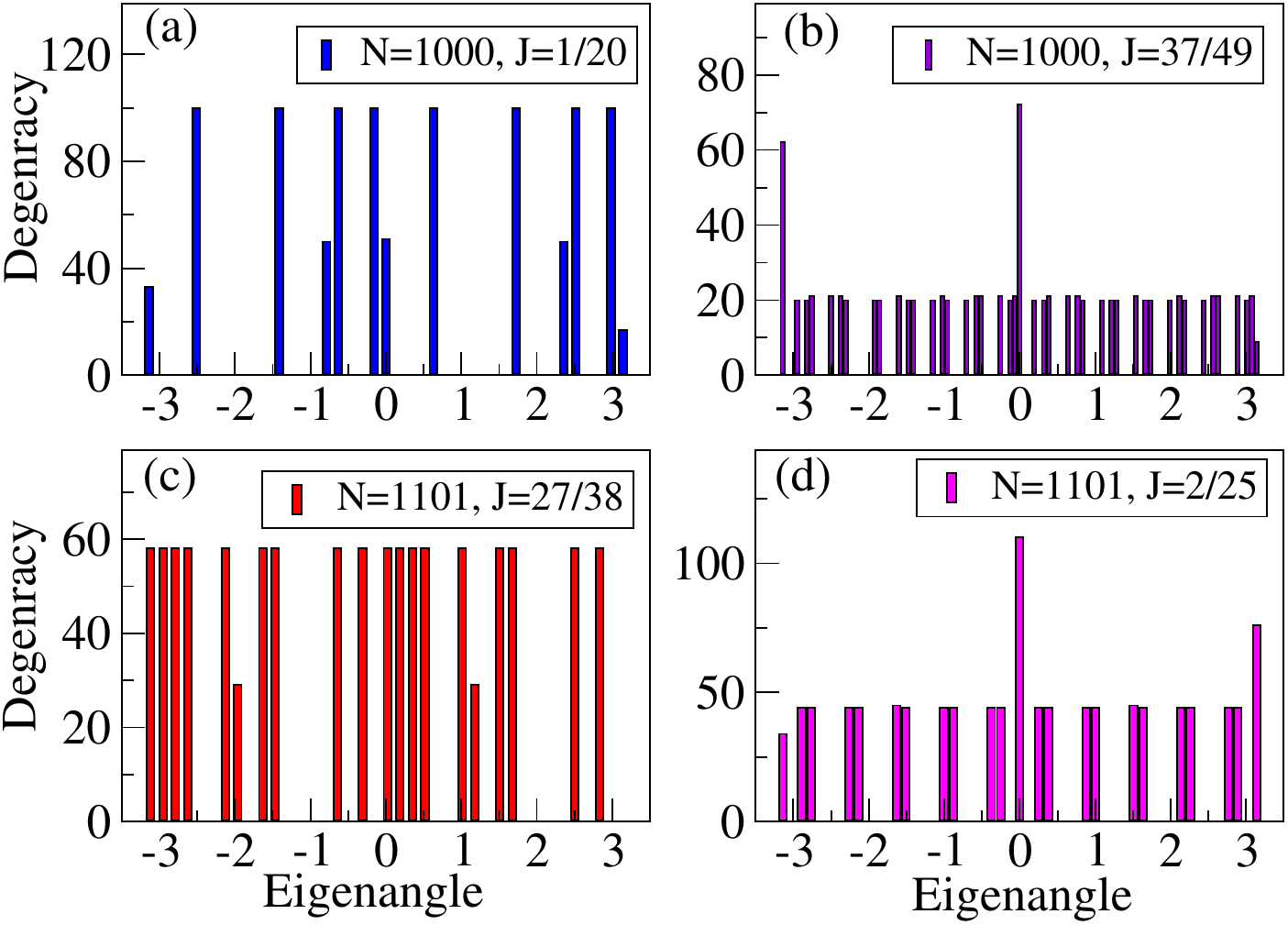}
\caption{Degeneracy of the eigenvalues of $\mathcal{U}$ for the rational values of $J$, $k'=J j^{p-1} \pi$ and $p=4$.}
\label{fig:degenrated}
\end{figure}

We use a quantity $\delta(n)=(\sum_{p,q}|\mathcal{U}_{p,q}^n-\mathcal{U}_{p,q}|)/2N$ to quantify the periodic nature of time-evolved unitary operator \cite{sharma2024exactly,sharma2024signatures}. If this quantity is zero, it implies the operator is  periodic. Using this quantity, we observe that, in Case~I~(a), the unitary operator exhibits periodicity for even $N$ and even $p$, with a period of $2ph$, as shown in Figs. \ref{fig:degenrated2111}~(a) and  \ref{fig:degenrated2111}~(b). For any odd-$N$ and even-$p$, the corresponding period is $ 2^{p+1} p  h$, as shown in Figs. \ref{fig:degenrated2111}~(c) and  \ref{fig:degenrated2111}~(d). Furthermore, we observe a high degree of spectral degeneracy in this case for both even and odd $N$ with even $p$, as shown in Fig. \ref{fig:degenrated}.  We also find that the unitary operator is periodic and its spectrum is highly degenerate in Case~II~(a) as well. However, in this case, the period depends on $N$, $h$, and $p$.  In contrast, we observed that for cases~I~(b) and ~II~(b), the eigenvalues are nondegenerate and the periodic behavior of the entanglement dynamics and operator  are absent   for any $N$ and even-$p$, as shown in Fig. \ref{fig:spectrum35}~ b and d.

Recent studies have shown that the presence of highly degenerated spectra, periodicity of entanglement dynamics, and unitary operator are the signatures of QI \cite{sharma2024exactly,sharma2024signatures}. This suggests QI for cases~I~(a) and ~II~(a), for any $N$ and even-$p$. In contrast, the aforementioned signatures of integrability are absent in cases~I~(b) and II~(b). However, the absence of these signatures does not necessarily imply that the system is non-integrable. To further characterize the nature of the system, we examine additional signatures of integrability, including the SFF, EE, higher-order spacing distributions, and spacing-ratio statistics, in Sec.~\ref{sec:example-section48}.

\section{For odd $p$-body interaction}\label{sec:example-section49}
In this section, we extend the analysis to the case  odd-$p$ and any $N,k'$ and arbitrary initial state. The rotational symmetries around y-axis is broken for odd-$p$ values and only the time-reversal symmetry remains \cite{munoz2021nonlinear}. Consequently, the Floquet operator can not be written in block-diagonlised structure. We will consider the even-$N$ and odd-$N$ cases separately.
\subsection{For even qubits}
The Floquet operator $\mathcal{U}$ for any even-$N$  can be written as,
\begin{eqnarray}
\begin{aligned}
\mathcal{U}
&=
\sum_{q=0}^{\frac{N}{2}}
i^{N}\cos(\theta_q)\,|q\rangle\langle q|  +
\sum_{q=0}^{\frac{N-2}{2}}
i^{N-2}\cos(\theta_q)\,
|q+\Delta\rangle\langle q+\Delta| \\
&\quad +
\sum_{q=0}^{\frac{N-2}{2}}
\sin(\theta_q)
\Big(
i^{N-1}|q\rangle\langle q+\Delta|
+
i^{N-3}|q+\Delta\rangle\langle q|
\Big),
\end{aligned}
\end{eqnarray}
where,
\[
\Delta=\frac{N+2}{2}, ~~~
s_1(q)=\frac{(N-2q)^p}{2^p}, ~~~ \mbox{and} ~~~
\theta_q=\frac{k'~s_1(q)}{p\,j^{\,p-1}}\,.
\]

 \begin{widetext}
The bra-ket notation provides a more compact description of the operator. It can be thus seen that, the matrix representation of the operator $\mathcal{U}$
consists of four blocks, each being diagonal. This description also eases the calculation of its time evolution. The $n$th time evolution of $\mathcal{U}$ can therefore be written as follows:
\begin{eqnarray}
\nonumber
\mathcal{U}^n
&=&\sum_{q=0}^{\frac{N}{2}}
\left[
\cos^2\!\left(\frac{n\pi}{2}\right)
+ i^{N}\cos(\theta_q)\sin^2\!\left(\frac{n\pi}{2}\right)
\right]
|q\rangle\langle q|+
\sum_{q=0}^{\frac{N-2}{2}}
\left[
\cos^2\!\left(\frac{n\pi}{2}\right)
+ i^{N-2}\cos(\theta_q)\sin^2\!\left(\frac{n\pi}{2}\right)
\right]
|q+\Delta\rangle\langle q+\Delta| \\
&&
+
\sum_{q=0}^{\frac{N-2}{2}}
\sin(\theta_q)\sin^2\!\left(\frac{n\pi}{2}\right)
\left(
i^{N-1}|q\rangle\langle q+\Delta|
+i^{N-3}
|q+\Delta\rangle\langle q|
\right).
\label{Eq:arbitaray130}
\end{eqnarray}
\end{widetext}
Substituting $n=2$ in Eq.~(\ref{Eq:arbitaray130}), we obtain $\mathcal{U}^2 = \mathcal{I}$. This result is independent of the parameter $k'$, implying that the eigenvalues of $\mathcal{U}$ belong to the set $\{ 1, -1 \}$ for any $k'$. The state $\ket{\psi_n}$ can be obtained by the $n$th implementation of the unitary operator $\mathcal{U}$ on the arbitrary initial state $\ket{\psi}$, expressed as:
\begin{eqnarray}
 \ket{\psi_n}&=&\mathcal{U}^n \ket{\psi}\\ \nonumber
 &=&\sum_{q=0}^{N/2-1}\left(B^+_q\ket{\phi_{q}^+}+B^-_q\ket{\phi_{q}^-}\right)+B^+_\frac{N}{2}\ket{\phi_{\frac{N}{2}}^+},
\end{eqnarray}
where the coefficients $B^+_q$, $B^-_q$ and $B^+_{{N}/{2}}$ can be  calculated by substituting the coefficients $a_{q+1}, b_{q+1}$ and $a_{\frac{N+2}{2}}$ from the Eqs. (\ref{Eq:arbitaray11}), (\ref{Eq:arbitaray12}) and (\ref{Eq:arbitaray14}) as:
\begin{eqnarray}\nonumber
 B^+_q&=&\sum^{\frac{N-2}{2}}_{q=0}\left(\mathcal{U}^n_{q,q}a_{q+1}+\mathcal{U}^n_{q,q+\Delta}b_{q+1}\right)\Big{/}\sqrt{2},\\ \nonumber
  B^-_q&=&\sum^{\frac{N-2}{2}}_{q=0}\left(\mathcal{U}^n_{q+\Delta,q+\Delta}b_{q+1}+\mathcal{U}^n_{q+\Delta,q}a_{q+1}\right)\Big{/}\sqrt{2}~~~~\mbox{and}\\ \nonumber
   B^+_{\frac{N}{2}}&=& \mathcal{U}^n_{\frac{N}{2},\frac{N}{2}}~a_{\frac{N+2}{2}}.\\\nonumber
\end{eqnarray}
The single-qubit RDM is then given as follows:
\begin{equation}
\rho_1(n)=\frac{1}{2}\left(
\begin{array}{cc}
 r_n & \bar{w}_n \\
\bar{w}_n^* & 2-r_n \\
\end{array}
\right),
\end{equation}
The coefficient $r_n $ and $\bar{w}_n$ can be calculated using the values of $B^+_q, B^-_q$ and $B^+_{\frac{N}{2}}$ in Eqs. (\ref{Eq:arbitaray125}) and (\ref{Eq:arbitaray126}). The eigenvalues $\lambda_1$ and $\lambda_2$ of single qubit, $\rho_1(n)$ are $\frac{1}{2} \left(1\pm\sqrt{1-{r}_n\left(2-{r}_n\right)-|\bar{{w}}_n}|^2\right)$. The linear entropy of single-qubit RDM is given as follows:
\begin{equation}\label{EQ:Entropy}
 S_{(\theta_0,\phi_0)}^{(N)}(n,k',p)=[r_n(2-r_n)-|{\bar{w}_n}|^2]/2.
\end{equation}
Using Eq.~\eqref{EQ:Entropy}, we can calculate the linear entropy  for  any $n,k'$, odd-$p$, even-$N$ and for an arbitrary initial state. We have shown that, for even $n$, the unitary operator becomes  identity, implying $|\psi_{2n}\rangle=|\psi\rangle$. Thus, the evolved state returns to the initial state. Consequently, an initially separable product state remains separable, and the linear entropy, given by Eq.~\eqref{EQ:Entropy}, is therefore zero for even $n$. This is true  for any $k'$, odd-$p$ and even-$N$. Hence, both the entanglement dynamics and the Floquet operator exhibit periodicity with period $2$. Thus, the system shows signatures of QI.
\subsection{For odd qubits}
The unitary operator $\mathcal{U}$ for this case  can be written as follows:
\begin{eqnarray}
\begin{aligned}
\mathcal{U}
&= \sum_{q=0}^{\frac{N-1}{2}} \Big\lbrace
\cos(\theta_q)
\Big(
i^{\,N-2}\,|q\rangle\langle q|
+
i^{\,N}\,|q+\Delta\rangle\langle q+\Delta|
\Big)
\\ &
+\sin(\theta_q)
\Big[
(-1)^{\frac{N+1}{2}} |q\rangle\langle q+\Delta|
+
(-1)^{\frac{N-1}{2}} |q+\Delta\rangle\langle q|
\Big] \Big\rbrace,
\end{aligned}
\end{eqnarray}
where, $\Delta={N+1}/{2}$. The $n$th time evolution of $\mathcal{U}$  can be written as,
\begin{eqnarray}\label{Eq:arbitaray131}
\begin{aligned}
\mathcal{U}^n
&= \sum_{q=0}^{\frac{N-1}{2}}\Big\lbrace \cos \left(\frac{n \pi}{2}\right)+
\cos(\theta_q) \sin \left(\frac{n \pi}{2}\right)
\Big(
i^{\,N-2}\,|q\rangle\langle q|
 \\ & +
i^{\,N}\,|q+\Delta\rangle\langle q+\Delta|
\Big)
 +
\Big[
(-1)^{\frac{N+1}{2}} |q\rangle\langle q+\Delta|
\\ & +
(-1)^{\frac{N-1}{2}} |q+\Delta\rangle\langle q|
\Big]\sin(\theta_q)\sin \left(\frac{n \pi}{2}\right)\Big\rbrace.
\label{Eq:arbitaray131}
\end{aligned}
\end{eqnarray}
The state $\ket{\psi_n}$ can be obtained by the $n$ successive application of the unitary operator $\mathcal{U}$ on the arbitrary initial state $\ket{\psi}$, expressed as,
\begin{eqnarray}
 \ket{\psi_n}&=&\mathcal{U}^n \ket{\psi}=\sum_{q=0}^{\frac{N-1}{2}}\left(B^+_q\ket{\phi_{q}^+}+B^-_q\ket{\phi_{q}^-}\right),
\end{eqnarray}
where the coefficients $B^+_q$ and  $B^-_q$ can be written as,
\begin{eqnarray}
 B^+_q&=&\sum^{\frac{N-1}{2}}_{q=0}\left(\mathcal{U}^n_{q,q}\tilde{a}_{q+1}+\mathcal{U}^n_{q,q+\Delta}\tilde{b}_{q+1}\right)\Big{/}\sqrt{2}~~\mbox{and}\\
  B^-_q&=&\sum^{\frac{N-1}{2}}_{q=0}\left(\mathcal{U}^n_{q+\Delta,q+\Delta}\tilde{b}_{q+1}+\mathcal{U}^n_{q+\Delta,q}\tilde{a}_{q+1}\right)\Big{/}\sqrt{2}.\\ \nonumber
\end{eqnarray}
The single-qubit RDM is then given as follows:
\begin{equation}
\rho_1(n)=\frac{1}{2}\left(
\begin{array}{cc}
 t_n & \bar{m}_n \\
\bar{m}_n^* & 2-t_n \\
\end{array}
\right),
\end{equation}
The coefficient $t_n $ and $\bar{m}_n$ can be calculated using the values of $B^+_q$ and $B^-_q$ in Eqs. (\ref{Eq:arbitaray129}) and (\ref{Eq:arbitaray128}).  The eigenvalues $\lambda_1$ and $\lambda_2$ of single qubit, $\rho_1(n)$ are $\frac{1}{2} \left(1\pm\sqrt{1-{t}_n\left(2-{t}_n\right)-|\bar{{m}}_n}|^2\right)$. The linear entropy of single-qubit RDM is given as follows:
\begin{equation}\label{EQ:Entropy1}
 S_{(\theta_0,\phi_0)}^{(N)}(n,k',p)=[t_n(2-t_n)-|{\bar{m}_n}|^2]/2.
\end{equation}
Substituting $n=2$ and $n=4$ in Eq.~(\ref{Eq:arbitaray131}), we obtain $\mathcal{U}^2 = -\mathcal{I}$ and $\mathcal{U}^4 = \mathcal{I}$. This result is independent of the parameter $k'$, and the eigenvalues of $\mathcal{U}$ belong to the set $\{ i, -i \}$ for all $k'$.
Thus, $\ket{\psi_{2n}} = -\ket{\psi}$, which differs from the initial state only by a global phase. Therefore,  the entanglement entropy remains the same for even $n$. If the initial state is a product state, the single-qubit RDM remains pure for all even $n$. Consequently, the linear entropy given in Eq. (\ref{EQ:Entropy1}), remains zero for all even $n$, for any $k'$ and odd $p$ with odd $N$. Hence, the entanglement dynamics exhibits periodicity with period $2$,
while the Floquet operator itself has periodicity $4$. The spectrum of $\mathcal{U}$ consists of only two distinct eigenvalues: $\pm 1$ for even $N$, and $\pm i$ for odd $N$. In both cases, these eigenvalues occur with equal degeneracy for any $k'$, and odd $p$. As we mentioned above that the presence of highly degenerated spectra, periodicity of entanglement dynamics, and unitary operator are the signatures of QI \cite{sharma2024exactly,sharma2024signatures}. Thus, the system is QI for any odd-$p$, $N$ and $k'$.

We have shown that signatures of QI, including highly degenerate spectra and periodicity of the entanglement dynamics and the unitary operator, are present in Cases~I~(a) and II~(a), but absent in Cases~I~(b) and II~(b), for arbitrary $N$ and even $p$. A similar situation is observed in our model for $p=2$, $\alpha=2$, and $k'=0.1$ \cite{haake1987classical}. In this case, the aforementioned signatures of integrability are absent; however, the level-spacing and spacing-ratio distributions follow Poisson statistics, which are  associated with integrable systems \cite{haake1987classical}. This observation demonstrates that the absence of the aforementioned signatures does not necessarily imply non-integrability, and that the system may still exhibit integrable behavior. Therefore, to characterize the nature of the system more conclusively for arbitrary $N$ and $p$, we perform a detailed analysis using additional spectral and entanglement-based indicators. This analysis is particularly focused on when $k'$ is not a rational multiple of $\pi$ (cases~I~(b) and II~(b)). Specifically, we examine the spectral statistics of the eigenvalues, the average adjacent gap ratio, the SFF, and the normalized average EEE ($\langle S\rangle/S\_{\mathrm{Max}}$) in the thermodynamic limit. The behavior of these quantities for the aforementioned parameter regimes will be discussed in the following sections. Throughout these sections, we consider values of $k'$ that are not rational multiples of $\pi$ (cases~I~(b) and II~(b)).

\section{EIGENVALUE STATISTICS}\label{sec:example-section48}
The eigenvalue statistics have emerged as a valuable tool in physics and used to study the integrable and chaotic nature of the system \cite{guhr1998random,mehta1963statistical,bohigas1991random,atas2013distribution,pandey2019quantum,Forrester2010}.  After resolving the symmetries in the spectra, the nearest-level statistics are used to determine the integrable or chaotic nature of a
system \cite{mehta1963statistical,berry1977level,bohigas1991random,atas2013distribution,pandey2019quantum}. On the other hand, long-range correlations are typically described by the Dyson-Mehta statistics $\Delta_3$ or the number variance $\Sigma^2$ \cite{jain1975higher,brody1981random,mitchell2009missing,bialous2016long,casal2021accuracy}. However, they are highly sensitive to the unfolding strategy, which can often lead to misleading results \cite{gomez2002misleading}. Alternatively, studying higher-order level spacings and gap ratios offers a more straightforward approach that is efficient both analytically and numerically \cite{rao2020higher,tekur2020symmetry,pandey2019quantum,Harshini2018a,bhosale2021superposition,bhosale2023universal,UdaysinhBhosaleScaling2018,rout2025higher}. In this section, we study the higher-order level spacings and higher-order spacing ratios  for  any $N$, even-$p$ for the cases~I~(b) and II~(b). Based on these statistics,  we conclude  that our model indeed shows QI.

\subsection{Nearest-Neighbor and Higher-Order Spacing Distributions}
In the non-integrable/chaotic cases, the level spacing follows the Wigner-Dyson distribution, implying level repulsion \cite{bohigas1984spectral,bohigas1984characterization}. In contrast, the integrable systems follow Poisson statistics, implying level clustering \cite{berry1977level,berry1984semiclassical}. These two statistics play a major role in differentiating the integrable and non-integrable systems. We will first apply the nearest level spacing analysis to our model, for the parameter $\alpha=\pi$.
\begin{figure}[t]\vspace{0.4cm}
\includegraphics[width=0.47\textwidth]{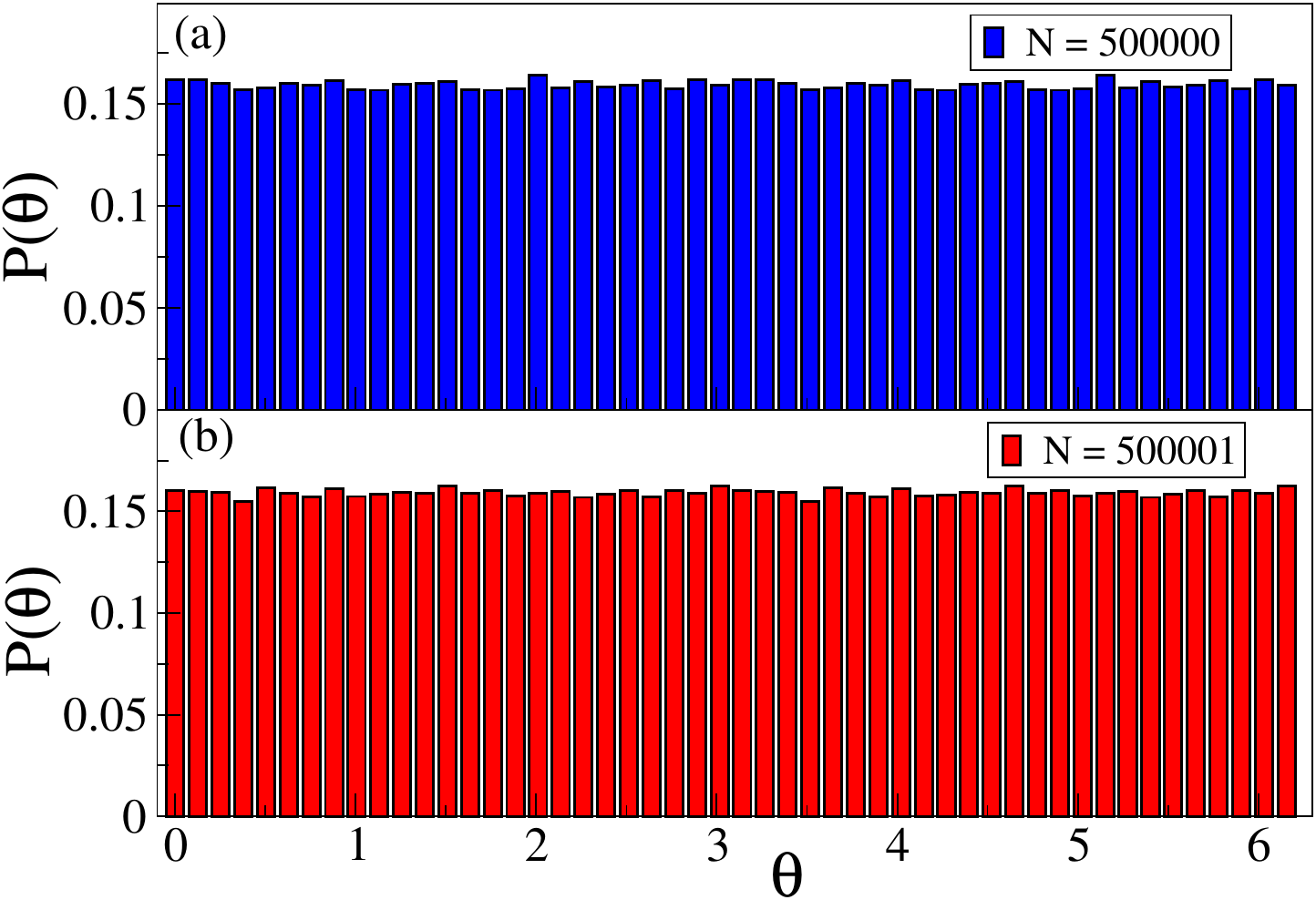}
\caption{The eigenvalue distribution for  $k^{\prime}=10$ and $p=2$, in the positive parity basis.}
\label{fig:degenrated1}
\end{figure}
For large $N$, around $5,00,000$, we observe that the eigenvalue distribution is nearly uniform for both even and odd-$N$, as shown in Fig.  \ref{fig:degenrated1}. After the unfolding process and resolving symmetries, we calculate the nearest level spacing $E_{i+1}-E_{i}$ and found that it follows the Poisson distribution ($\exp(-s)$) for even and odd-$N$. The results are plotted in  Fig. \ref{fig:degenrated2}. In contrast, even if a system is quantum chaotic, the NN statistics may sometimes give incorrect results due to the presence of symmetries, even for the presence of a single symmetry \cite{KarolTensor2012,giraud2022probing}. In this case, if there is no prior knowledge of the symmetry structure and the Hamiltonian is considered  as a whole, then NN spectral statistics gives the Poisson distribution \cite{bhosale2021superposition,pandey2019quantum,tkocz2012tensor,giraud2022probing}. Thus, knowledge of symmetries and resolving them becomes necessary in this case. Even if some hidden symmetries are present, the higher-order statistics can be used to distinguish the integrable or chaotic nature correctly \cite{bhosale2021superposition,tekur2020symmetry}.

The higher-order spacing  and higher-order spacing ratio distributions \cite{tekur2020symmetry,bhosale2021superposition} are used to distinguish between the integrable and non-integrable nature of the systems, specially when symmetries are present. The higher-order level spacing are defined as:
\begin{equation}
 s_{i}^{(k)}=E_{i+k}-E_{i},~~~~~i,k=1,2,3,\dots.
\end{equation}
For integrable system, i.e., when the spectra is uncorrelated, its distributions are given as follows \cite{rao2020higher}:
\begin{equation}\label{Eq:Poisson}
 P_P^{(k)}(s)=\frac{k^k}{(k-1)!}s^{k-1} \exp(-k s).
\end{equation}
We found that our results agree very well with Eq.~(\ref{Eq:Poisson}).
The results are plotted  in Fig. \ref{fig:degenrated2} for $k=2$ to $4$. We observe that it also holds true for  $k\ge 5$, cases as well (results are not shown here). Thus, this implies that our system is QI for these parameter values.

\begin{figure}[t]\vspace{0.4cm}
\includegraphics[width=0.47\textwidth]{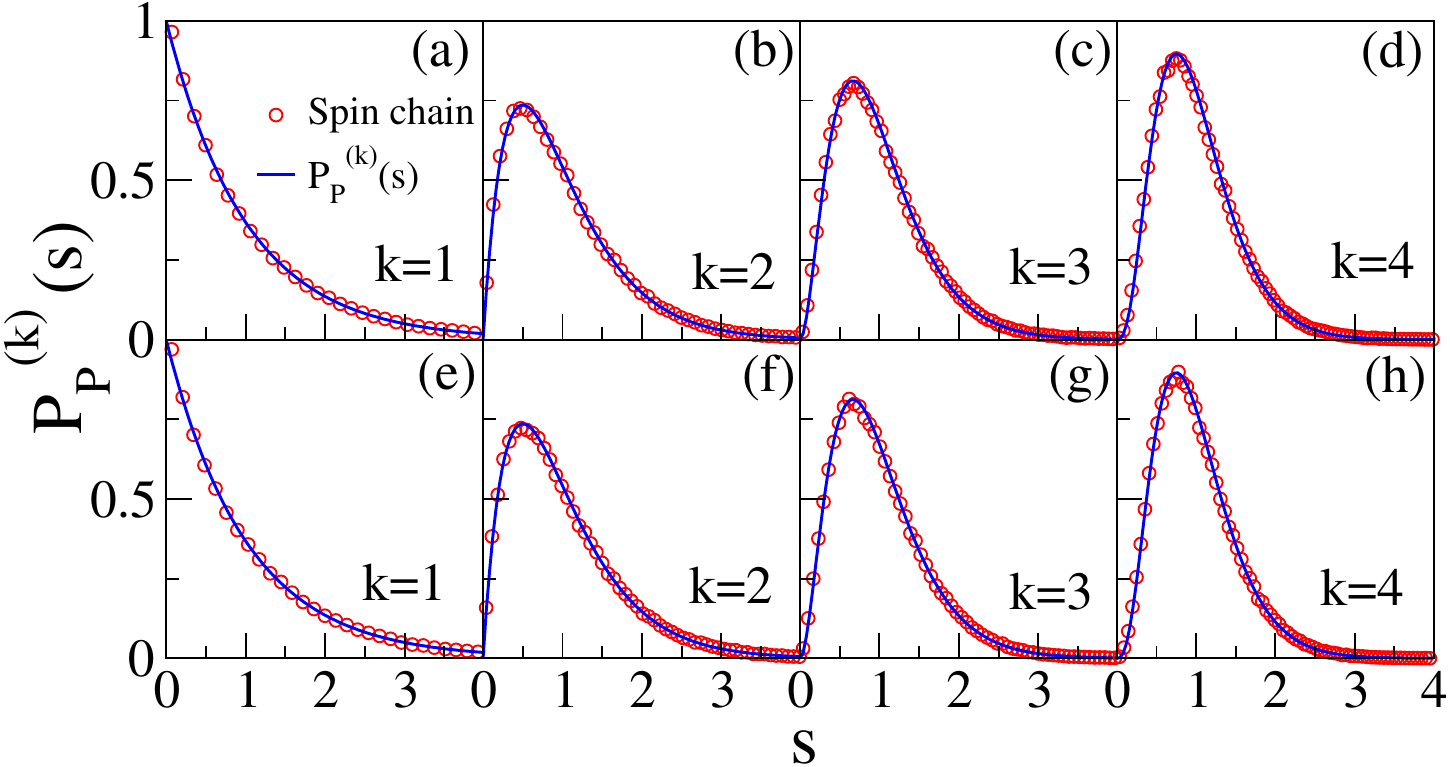}
\caption{Higher-order level spacing  distributions for $k = 1$ to $4$. Here $N=500000$ (upper panel) and $N=500001$ (lower panel) for the parameters $k'=10$ and $p=2$, in the positive parity basis.}
\label{fig:degenrated2}
\end{figure}
\begin{figure}[t]\vspace{0.4cm}
\includegraphics[width=0.47\textwidth]{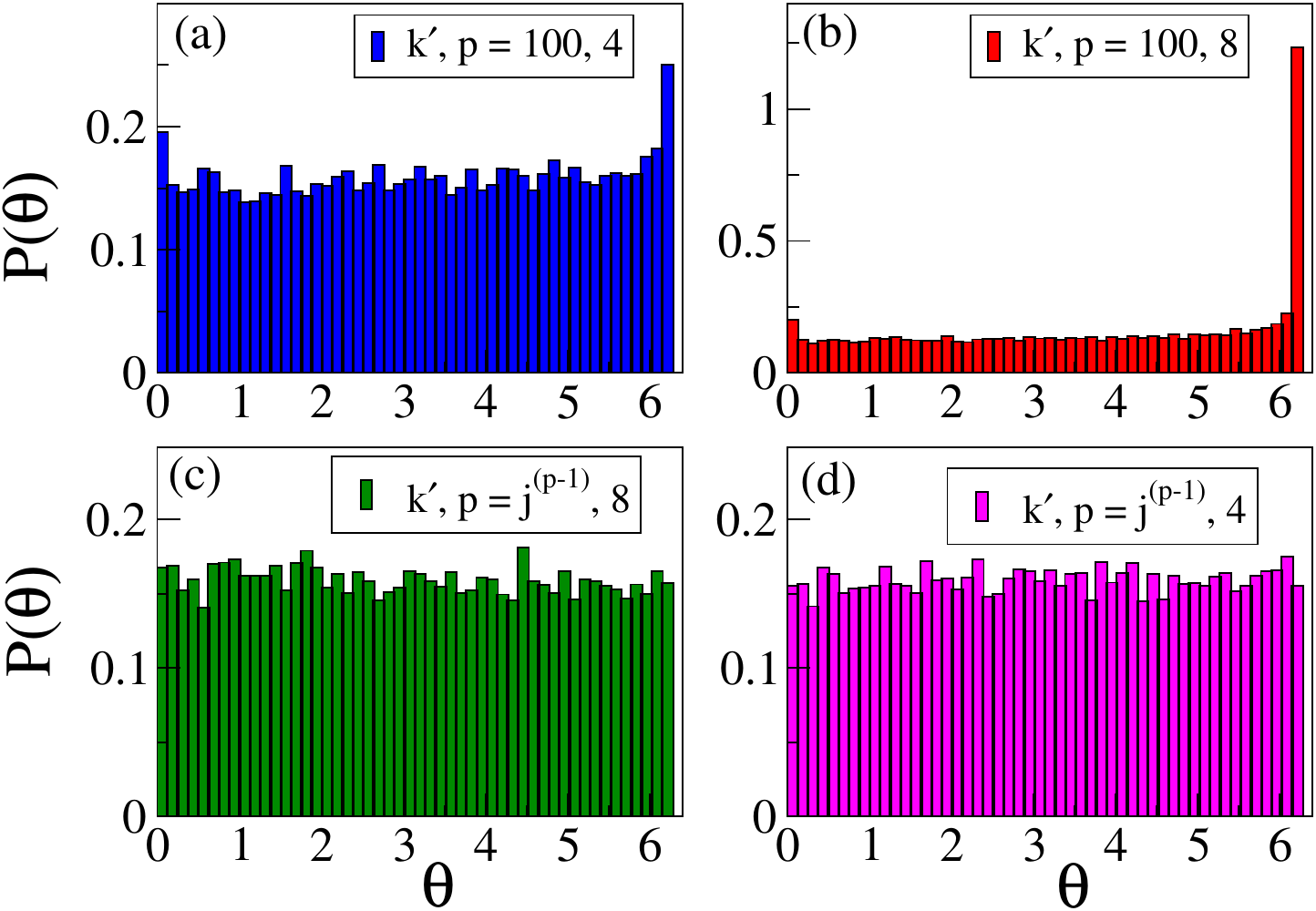}
\caption{The eigenphase distributions for various values of $k^{\prime}$ and $p$ with  $N = 40001$~($j=20000.5$), in the positive parity basis.}
\label{fig:degenrated12}
\end{figure}
\begin{figure}[t]\vspace{0.4cm}
\includegraphics[width=0.47\textwidth]{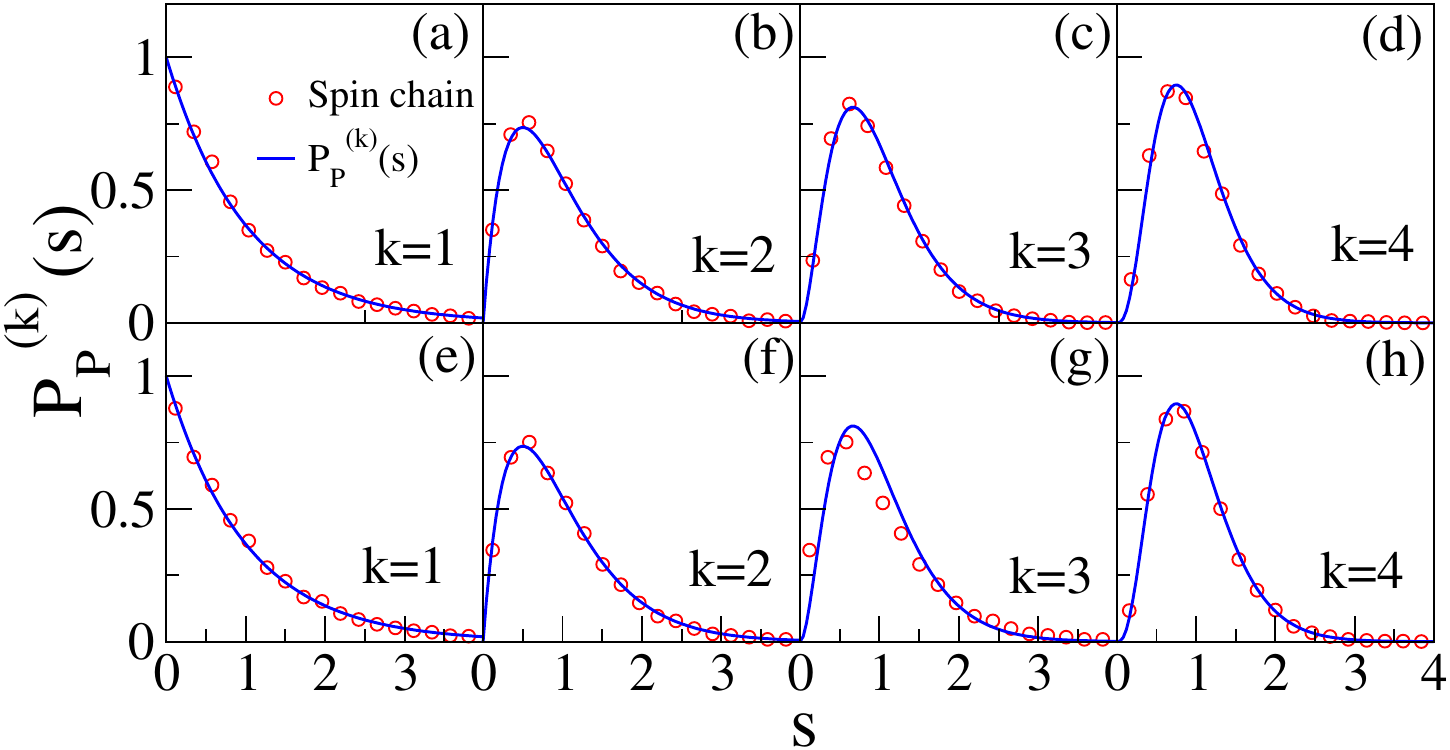}
\caption{Higher-order level spacing distributions of the bulk region for $k = 1$ to $4$ at $p = 8$ (upper panel) and $p = 20$ (lower panel), with $N = 40001$ and $k^{\prime} = 100$, in the positive parity basis after removing the spike contributions.}
\label{fig:degenrated21}
\end{figure}
As the nonlinearity parameter \(p\) is increases, the eigenvalue distribution becomes non-uniform. For larger values \(p > 2\), the eigenphases tend to accumulate within  the small regions close to \(2\pi\). This leads to pronounced spikes in the distribution of eigenphases, particularly for small \(k'\), as shown in Figs.~\ref{fig:degenrated12}~(a) and \ref{fig:degenrated12}~(b). However, we observe that while keeping \(p\) fixed, the extent of the spiky region gradually decreases with $k’$. In the limit of large \(k'\), specifically when \(k' = j^{(p-1)}\), the eigenvalue distribution becomes uniform for any value of \(p\), as illustrated in Figs.~\ref{fig:degenrated12}~(c) and \ref{fig:degenrated12}~(d).

\begin{figure}[t]\vspace{0.4cm}
\includegraphics[width=0.47\textwidth]{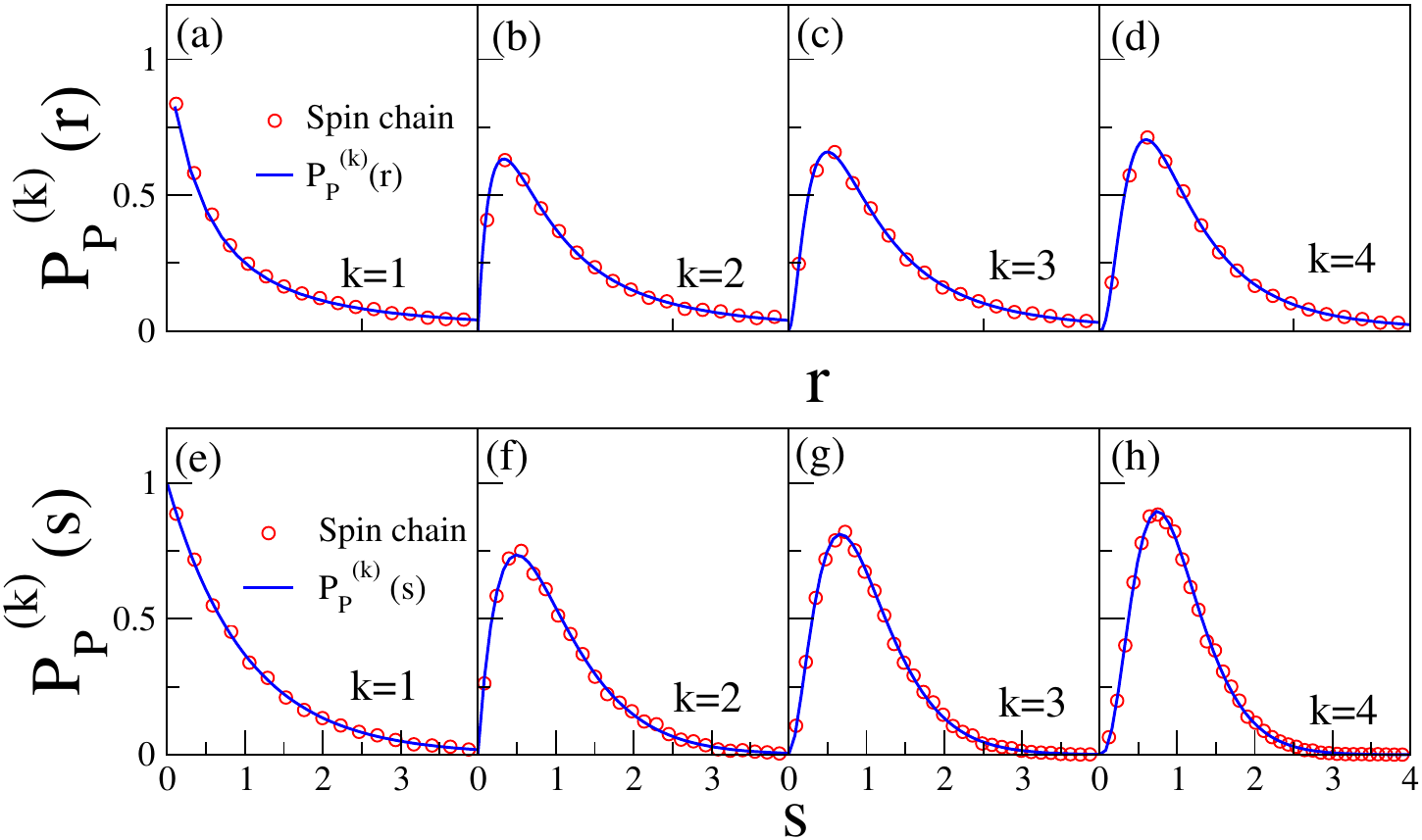}
\caption{Higher-order spacing ratio (upper panel) and spacing (lower panel) distributions for $k = 1$ to $4$, with $k^{\prime} = j^{p-1}$, $N = 40001$ and $p = 8$, in the positive parity basis.}
\label{fig:degenrated22}
\end{figure}

From Figs.~\ref{fig:degenrated12}~(a) and \ref{fig:degenrated12}~(b), it can be seen that the eigenphase distribution exhibits pronounced spikes near $0$ and $2\pi$, indicating strong eigenvalue accumulation in these regions. Apart from these small regions, the spectrum is approximately uniformly distributed \cite{jisha2024universality,schreiber1999energy,grimm2002energy}. Therefore, we restrict our analysis to this bulk region and perform spectral unfolding using standard techniques such as polynomial unfolding and local polynomial fitting (sliding window method) \cite{abuelenin2018spectral,abul2014unfolding,flores2001spectral}. After the unfolding process, we find that the results are in excellent agreement with the Poisson distribution, as given in Eq.~(\ref{Eq:Poisson}). The results are shown in Fig.~\ref{fig:degenrated21} for $k = 1$ to $4$. We have verified that this behavior persists for $k \geq 5$ as well (results are not shown here). We also observed the similer behaviour persist for large $k'=j^{p-1}$ as well, shown in lower panels of Fig.~\ref{fig:degenrated22}. It is important to note that the  spacing statistics are sensitive to the unfolding procedure, and improper unfolding can lead to misleading conclusions. Therefore, to validate  our results in both the bulk and non-uniform regions of the spectrum, we further employ statistical measures that are independent of unfolding. These will be discussed in the next subsection.
\subsection{Nearest-Neighbor and Higher-Order Spacing Ratio Distributions}
Another widely used quantity for studying spectral fluctuation in RMT is the spacing ratios \cite{oganesyan2007localization,atas2013distribution}. The advantage of this quantity is  that it is independent of the local DOS (density of states) and does not require an unfolding process.  The $k${th} order non-overlapping spacing ratio \cite{Harshini2018a,tekur2020symmetry,bhosale2021superposition,bhosale2023universal,UdaysinhBhosaleScaling2018} is  defined as,
\begin{align}
 r_i^{(k)} = \frac{s_{i+k}^{(k)}}{s_{i}^{(k)}} = \frac{E_{i+2k}-E_{i+k}}{E_{i+k}-E_i}, ~~~~~~~
i,k=1,2,3,\dots .
\label{hosr}
\end{align}
This ratio has been used to study higher-order fluctuation statistics in the Gaussian \cite{Harshini2018a,bhosale2023universal},
circular \cite{Harshini2018a,bhosale2021superposition}, and
Wishart ensembles \cite{UdaysinhBhosaleScaling2018}. We have already shown that the higher-order spacings follow the Poisson statistics. To further support our results, we study the higher-order spacing ratios distributions. If the system is integrable, then
the $k$-{th} order spacing ratio distribution is given by \cite{tekur2020symmetry},
\begin{align}
 P_P^{(k)}(r)&= \frac{(2k-1)!}{\big((k-1)!\big)^2}\frac{r^{k-1}}{(1+r)^{2k}}.
\end{align}

\begin{figure}[t!]\vspace{0.4cm}
\includegraphics[width=0.47\textwidth]{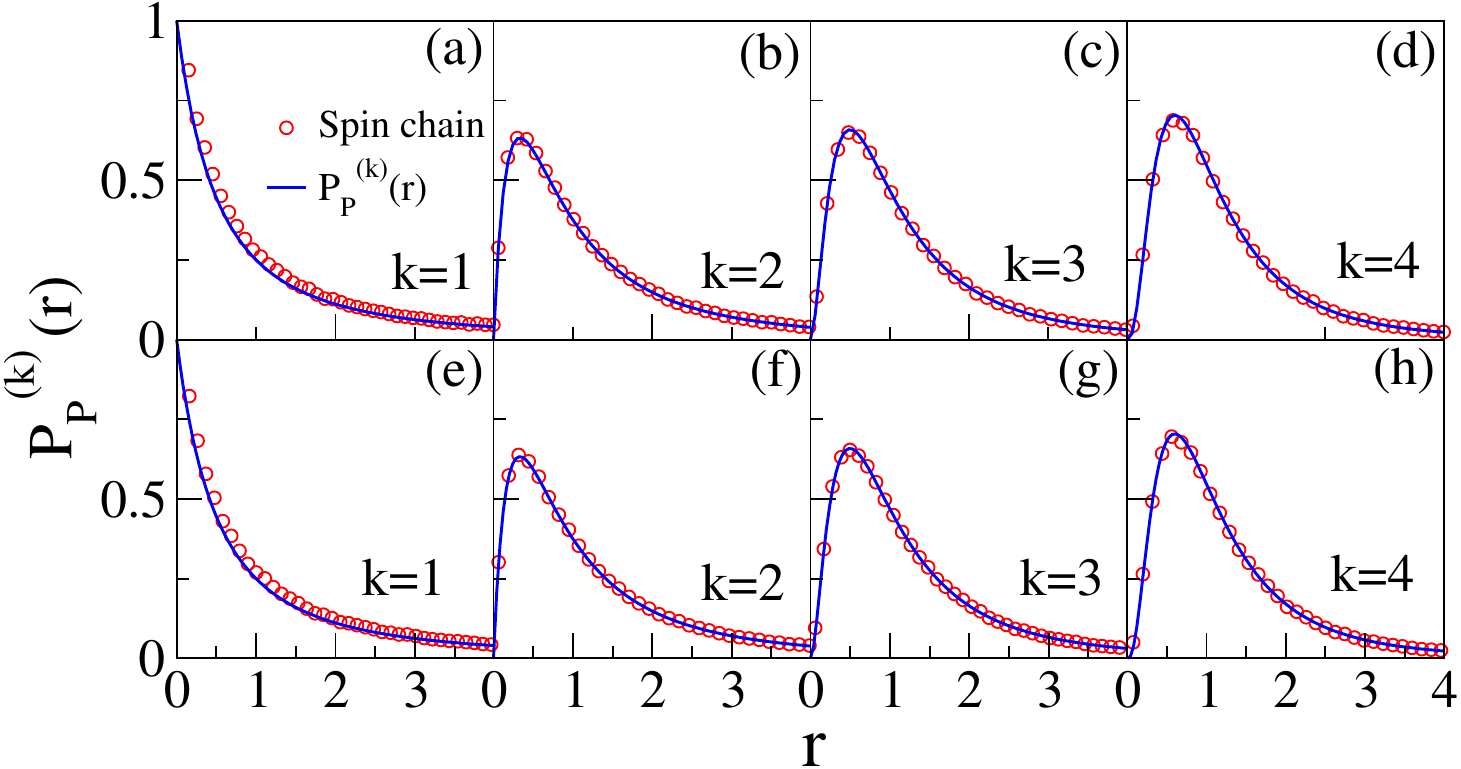}
\caption{Higher-order spacing ratio distributions for $k = 1$ to $4$. The upper (lower) panel corresponds to $N = 500000$ ($N = 500001$). The parameters are $k^{\prime} = 15$, and $p = 2$, in the positive parity sector.}
\label{fig:spectrum}
\end{figure}
For $k=1$, it reduces to the familiar form \cite{atas2013distribution,tekur2020symmetry},
\begin{equation}\label{eq:k=1}
 P_P^{(1)}(r)=\frac{1}{(1+r)^2}.
\end{equation}
For $k=2$,
\begin{equation}\label{eq:k=2}
P_P^{(2)}(r)=\frac{6r}{(1+r)^4},
\end{equation}
 for $k=3$,
\begin{equation}\label{eq:k=3}
 P_P^{(3)}(r)=\frac{30r^2}{(1+r)^6},
\end{equation}
and for $k=4$,
\begin{equation}\label{eq:k=4}
 P_P^{(4)}(r)=\frac{140r^3}{(1+r)^8}.
\end{equation}
\begin{figure}[t!]\vspace{0.4cm}
\includegraphics[width=0.47\textwidth]{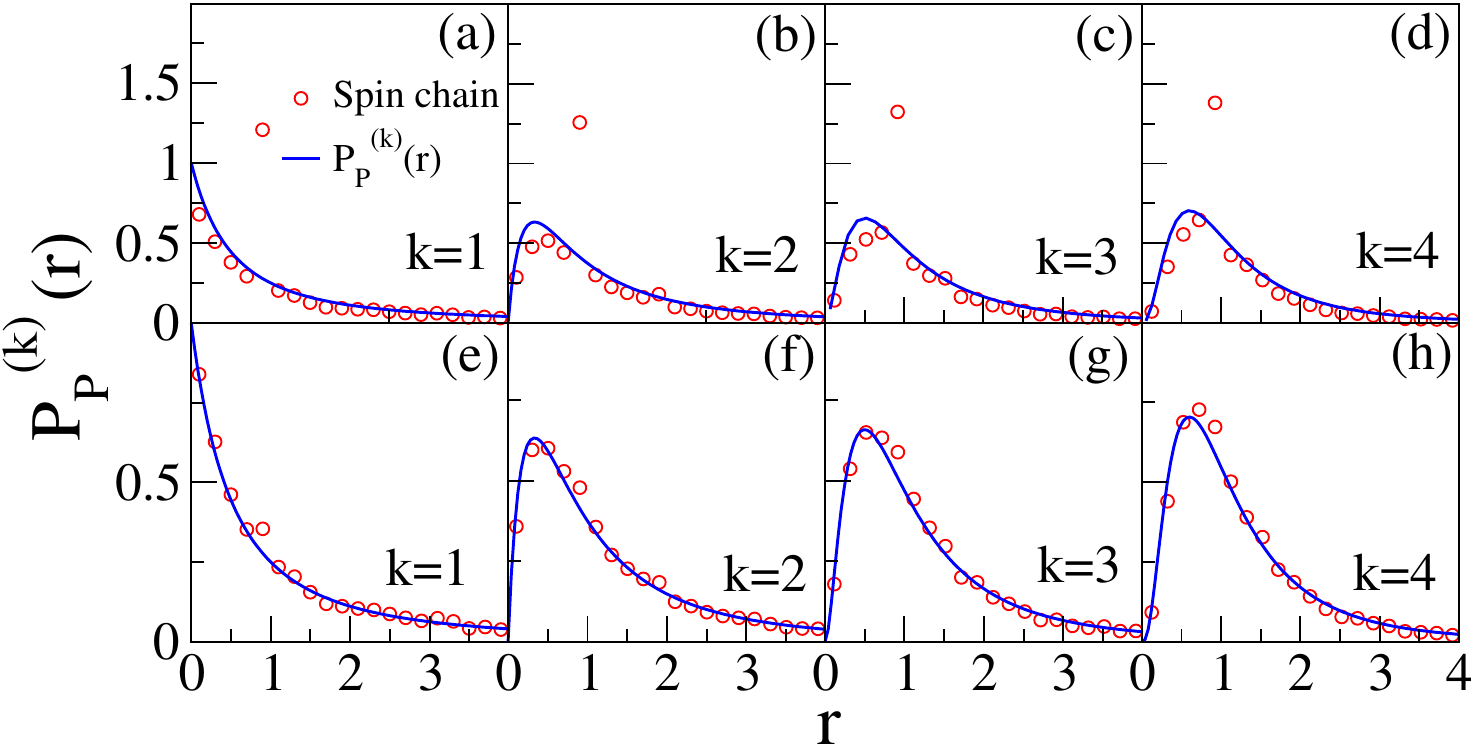}
\caption{Higher-order spacing-ratio distributions for $k = 1$ to $4$: full spectrum (upper panel) and after removal of quasi-degeneracy-induced spikes (lower panel) in the positive parity sector, with $N = 400001$, $k^{\prime} = 100$ and $p = 8$. The spike originate from a accumulation of  large eigenvalues within a narrow spectral region.}
\label{fig:spectrum11}
\end{figure}
Our results are in excellent agreement with Eqs.~(\ref{eq:k=1})-(\ref{eq:k=4}), as illustrated in Fig.~\ref{fig:spectrum}.
This confirms that the system follows Poisson statistics for the considered values of $k'$, any $N$ and $p=2$. As mentioned earlier, for even $p$, eigenphase accumulation occurs in specific regions of the spectrum. To characterize the behavior of the full spectrum, we employ ratio statistics, as shown in the upper panel of Fig.~\ref{fig:spectrum11}. We observe a pronounced peak around $r = 1$; however, apart from this feature, the distribution is in very close agreement with Poisson statistics for any $k'$ and even-$p$. In contrast, the lower panel of the same figure shows the bulk spectrum, where the density of states is nearly uniform. In this region, the distribution  of eigenvalues shows excellent agreement with Poisson statistics, and the peak near $r = 1$ disappears. We find that this behavior persists for $k \geq 5$ as well (results not shown here). Furthermore, as illustrated in the upper panels of  Fig.~\ref{fig:degenrated22}, these results persist for higher $k’=j^{p-1}$ as well. Thus, the eigenvalue statistics confirm the QI nature of the system for the considered values of  $k'$, even-$p$, and any $N$.

\subsection{Average Adjacent Gap Ratio}
\begin{figure}[t!]\vspace{0.4cm}
\includegraphics[width=0.47\textwidth]{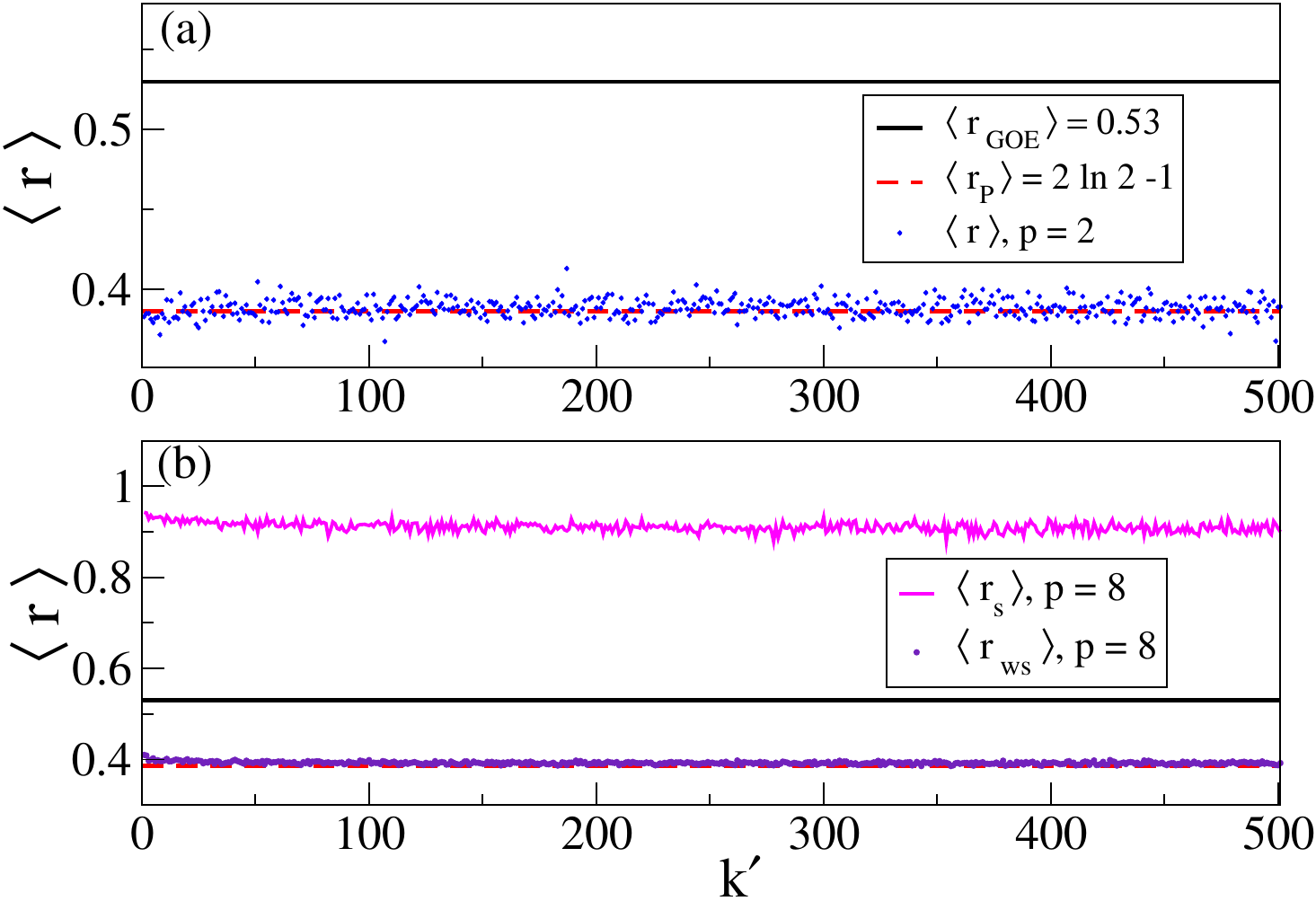}
\caption{Adjacent gap ratio $\langle r \rangle$ vs $k^{\prime}$ for (a) $N = 40000$, $p = 2$ and (b) $N = 40001$, $p = 8$, where panel (b) includes both spike-region ($\langle r_s \rangle$) and without spike region ($\langle r_{\mathrm{ws}} \rangle$) results in the positive parity subspace.}
\label{fig:spectrum42}
\end{figure}

Now, we move on to yet another important spectral signature, the  adjacent gap ratio \cite{oganesyan2007localization,atas2013distribution}. This quantity is useful  for detecting the integrable and chaotic nature of the system. The average adjacent spacing ratio  is defined as,
\begin{eqnarray}
 \langle r\rangle=\frac{1}{D-2}\sum_{j=1}^{D-2} r_j,~~~r_j=\frac{\mbox{min}(s_j,s_{j+1})}{\mbox{max}(s_j,s_{j+1})},
\end{eqnarray}
where $s_j=E_{j+1}-E_j$ is the spacing of the  consecutive eigenphases, and $D=N+1$ is the dimension of the Hilbert space. Initially, it was introduced to characterize the change
of statistics across a many-body localization (MBL) transition. The average ratio  is expressed as an ensemble average:
\begin{equation}
 \langle r \rangle =\int_0^1 rP(r)dr.
\end{equation}
 The exact analytical values of $\langle r \rangle$ for integrable systems is $\langle r \rangle_{{P}}=2~\log 2-1$ and for  quantum chaotic systems is
$\langle r \rangle_{{GOE}}=0.536$ \cite{atas2013distribution,giraud2022probing}. This quantity is particularly useful for tracking the transition from chaos to integrability as a function of a Hamiltonian parameter. We numerically study the behavior of $\langle r \rangle$  with $k'$ and $\alpha=\pi$. We observe that for the considered values of $k'$, $p=2$ and any $N$, the average adjacent ratio is consistent with $\langle r \rangle_{P}$. This is  shown in upper panel of the Fig. \ref{fig:spectrum42}. We have already shown that peaks in the density of states (DOS) of eigenphases arise around specific values that depend on $p$. In contrast, the higher-order spacing ratios and spacing distributions in the bulk region show excellent agreement with Poisson statistics. A similar behavior is also observed for the average adjacent gap ratio $\langle r \rangle$, which is consistent with the Poisson value for any $k'$ and even $p$. This can be seen in the lower panel of the Fig. \ref{fig:spectrum42}. On the other hand, in regions where the DOS is non-uniform or exhibits sharp peaks, the average adjacent gap ratio approaches $\sim 0.9$, as shown in the lower panel of Fig. \ref{fig:spectrum42}. This indicates the presence of strong quasi-degeneracies or hidden symmetries in that region of the spectrum.
 We further find that, for any $N$, even $p$, and $\alpha = m\pi$, the average higher-order gap ratios as a function of $k'$ for bulk eigenphases coincide with the corresponding Poisson values (results  are not shown here).

Based on the analysis of higher-order level spacing distributions, higher-order spacing ratios, and both the average adjacent and higher-order gap ratios as functions of $k’$, we conclude that the system exhibits Poisson statistics, a signature of integrability. Although nearest-neighbor level spacing statistics effectively characterize short-range correlations, they are insensitive to long-range spectral correlations. To gain deeper insight into the global structure of the spectrum, we therefore study the SFF in the following section.


\subsection{Spectral Form Factor}
The NN spacing and the average adjacent gap ratio probe only short-range, time-independent spectral correlations; they do not capture long-range or dynamical correlations \cite{Haakebook,Prasad:2025afd}. Therefore, to investigate the long-range correlations and time-dependent signatures beyond the local level statistics, global measures like Spectral Form Factor (SFF) and number variance are very useful. The SFF is  an important and simple measure of long-range spectral diagonstic that captures all the  eigenvalues of the spectrum at once. It is basically a Fourier transformation of a two-level correlation function, which is widely used as a tool in distinguishing the integrable and chaotic nature of the system \cite{brezin1997spectral,berry1985semiclassical,Haakebook}. The SFF and connected SFF (CSFF) \cite{Haakebook,dong2025measuring} are denoted as $K(t,N)$ and $K_c(t,N)$, respectively, and are defined as follows:
\begin{eqnarray}\nonumber
K(t,N)&=&\big \langle |\mbox{Tr}[\mathcal{\hat{U}}(t)]|^2 \big\rangle=\Big \langle \sum_{a,b}\exp \left(i \left(E_a-E_b\right)t\right) \Big\rangle,\\
K_c(t,N)&=&\left[K(t,N)-\Big|\Big \langle \sum_{a=1}\exp \left(i \left(E_a\right)t\right) \Big\rangle \Big|^2 \right],
\end{eqnarray}
\begin{figure}[t]\vspace{0.4cm}
\includegraphics[width=0.47\textwidth, height=0.35\textwidth]{SFF_plot.png}
\caption{Spectral form factor $\langle K(t) \rangle$ as a function of time $t$. The colored solid lines show numerical results, while the black dashed lines represent the Poisson prediction for $N=2000$ and $3000$ realizations. Panels correspond to (a) $k’=100, p=2$, (b) $k’=100,p=4$, (c) $k'=j^{(p-1)}, p=4$, and (d) $k'=\sqrt{2} j^{(p-1)}, p=8$. The behavior follows the Poisson form, indicating the absence of level correlations and presence of integrable nature.}
\label{fig:spectrum489}
\end{figure}
where $\big \langle \dots \big \rangle $ donates the ensemble average, $\lbrace E_a \rbrace$ are the quasienergies or eigenenergies of the $\mathcal{\hat{U}}$ and $\hat{H}$ respectively and $D$ is the dimension of Hilbert-space. The SFF has been extensively used in various areas of physics, such as  many-body quantum chaos, MBL systems, black holes, as well as in other related fields \cite{liu2018spectral,prakash2021universal,dong2025measuring}. The behaviour of SFF is basically divided into three distinct time domains: early time, intermediate time, and late time. Each time domain is important and provides distinct spectral and dynamical information. The early time behaviour of SFF, including dips and subsequent oscillations, is governed by a non-universal feature of the spectrum. These features  vary from system to system. Beyond the early-time regime, the SFF exhibits a linear ramp for quantum chaotic systems, while integrable systems show a power-law decay or no ramp between the Thouless time ($t_T$) and the Heisenberg time ($t_H$), followed by  a saturation at the late-time plateau \cite{Prasad:2025afd,Haakebook,dong2025measuring}. There time scales are defined subsequently.  The Thouless and Heisenberg times are  fundamental time scales  to understand the dynamics of complex systems, including  quantum chaotic systems, disordered systems, quantum systems out of equilibrium, and MBL systems \cite{schiulaz2019thouless,chan2018spectral,prakash2021universal,martinez2025decomposing}. The Thouless time  is crucial for understanding the transition from the localised phase to the delocalised phase, and  it is related to the diffusive behavior of a system.  The  interval between Thoulas time and Heisenberg time plays a key role in distinguishing the integrable and chaotic nature of the system. During this time interval, chaotic systems show a universal ramp feature in SFF \cite{mehta2004random,dong2025measuring,martinez2025decomposing}. In contrast, integrable and MBL systems lack this feature and typically exhibit non-universal  power law decay \cite{prakash2021universal,riser2020nonperturbative,riser2021power}.
The analytical expression of SFF for Poisson numbers is given as follows;
\begin{equation}\label{ch22}
K^P(t)=N+\frac{2}{(\mu t)^2}-\frac{(1+i \mu t)^{1-N}+(1-i \mu t)^{1-N}}{(\mu t)^2}.
\end{equation}
\begin{figure}[t]\vspace{0.4cm}
\includegraphics[width=0.47\textwidth, height=0.20\textwidth]{SFF_plot1.png}
\caption{Spectral form factor $\langle K(t) \rangle$ vs $t$ in the positive parity sector: unfolded without spike (ws) (blue) and spike-region (orange) spectra, compared with Poisson (black) and GOE (red). (a) $k^{\prime} = 1$, (b) $k^{\prime} = 100$, with $N = 40001$ and $p = 8$.}
\label{fig:spectrum4891}
\end{figure}
The quantum chaotic systems with time-reversal symmetry show RMT statistics described  by  GOE (or circular), for which the analytical expressions \cite{kos2018many,Haakebook} are given as follows:
\begin{align}
K^{\mathrm{GOE}}(t)
&= K^{\mathrm{GOE}}_{c}(t)
+ \left[
\frac{\pi}{\mu t}
J_1\!\left(\frac{2N\mu t}{\pi}\right)
\right]^2 ,
\label{ch21}
\\[6pt]
K^{\mathrm{GOE}}_{c}(t)
&= N
\begin{cases}
\displaystyle
\frac{\mu t}{\pi} - \frac{\mu t}{2\pi} \ln\!\left(1+\frac{\mu t}{\pi}\right),
& 0 < \mu t < 2\pi, \\[8pt]
\displaystyle
2 - \frac{\mu t}{2\pi} \ln\!\left(\frac{\mu t+\pi}{\mu t-\pi}\right),
& 2\pi < \mu t < \infty ,
\end{cases}
\label{eq:GOE_connected}
\end{align}
where $J_1(x)$ is the Bessel function of the first kind, $t_H = 2\pi / \mu$ is the Heisenberg time, and $t_T \approx 1/(\mu N)$ is the Thouless time, respectively and $\mu$ is mean level spacing. We calculate the SFF after unfolding the bulk spectrum of the eigenphase distribution. For any $k’$, the SFF for even-$p$ follows the analytical Poisson prediction given in Eq. (\ref{ch22}), indicating that the system is integrable. This behavior is   shown in Figs. \ref{fig:spectrum489} and \ref{fig:spectrum4891}. We observed that for odd $p$, the SFF exhibits strong fluctuations due to the presence of high degeneracy for any value of $k’$. Moreover, the absence of the characteristic ramp-plateau structure indicates the lack of level repulsion. The oscillatory behavior at short times gradually decays and approaches $\langle K(t) \rangle \approx 1$ at long times. In contrast, the spike region (without unfolding) exhibits non-universal features arising from the clustering of eigenphases within a narrow spectral window, leading to pronounced spikes in the DOS, shows power-law decay and absence of ramp. This can be seen in  Fig.~\ref{fig:spectrum4891}. Thus, this behavior suggests that the system exhibits quantum integrability for any $N$, even-$p$ and the considered values of $k'$.
\section{Average Eigenstate Entanglement Entropy} \label{sec:example-section4888}
In this section, we  compute the average EE of energy eigenstates. Previous studies  have shown that the average EE of energy eigenstates for  non-vanishing bipartition is a useful diagnostic for distinguishing integrable and non-integrable systems \cite{vidmar2017entanglement,hackl2019average,lydzba2020eigenstate,kumari2022eigenstate,page1993average}.
In Refs. \cite{vidmar2017entanglement,hackl2019average,lydzba2020eigenstate,kumari2022eigenstate,alba2015eigenstate,swietek2024eigenstate,vidmar2017entanglement,bianchi2022volume}, the typical integrable systems are studied.
There, it has been analytically shown that for a given bipartition ($N_a/2,(N-N_a)/2$), the ratio $\langle S \rangle/S_{Max}$  in the thermodynamic limit for free fermions lies within the interval $[0.52,0.59]$, for the XY chain  $0.557$  and $0.5$ for the LMG model. Whereas, in some of the integrable models, it has been observed that the average eigenstate EE follows volume-law scaling \cite{swietek2024eigenstate}.
Thus, for integrable models, the asymptotic value of the  ratio $\langle S \rangle/S_{Max}$, for non-vanishing partition, lies between $0$ and $1$. Whereas, for typical nonintegrable systems, it is close to $1$ \cite{page1993average,lakshminarayan2001entangling,seshadri2018tripartite}.
Thus, whenever the asymptotic value of the ratio is less than $1$, it can be concluded that the model is QI.

\begin{figure}[t]\vspace{0.4cm}
\includegraphics[width=0.47\textwidth, height=0.35\textwidth]{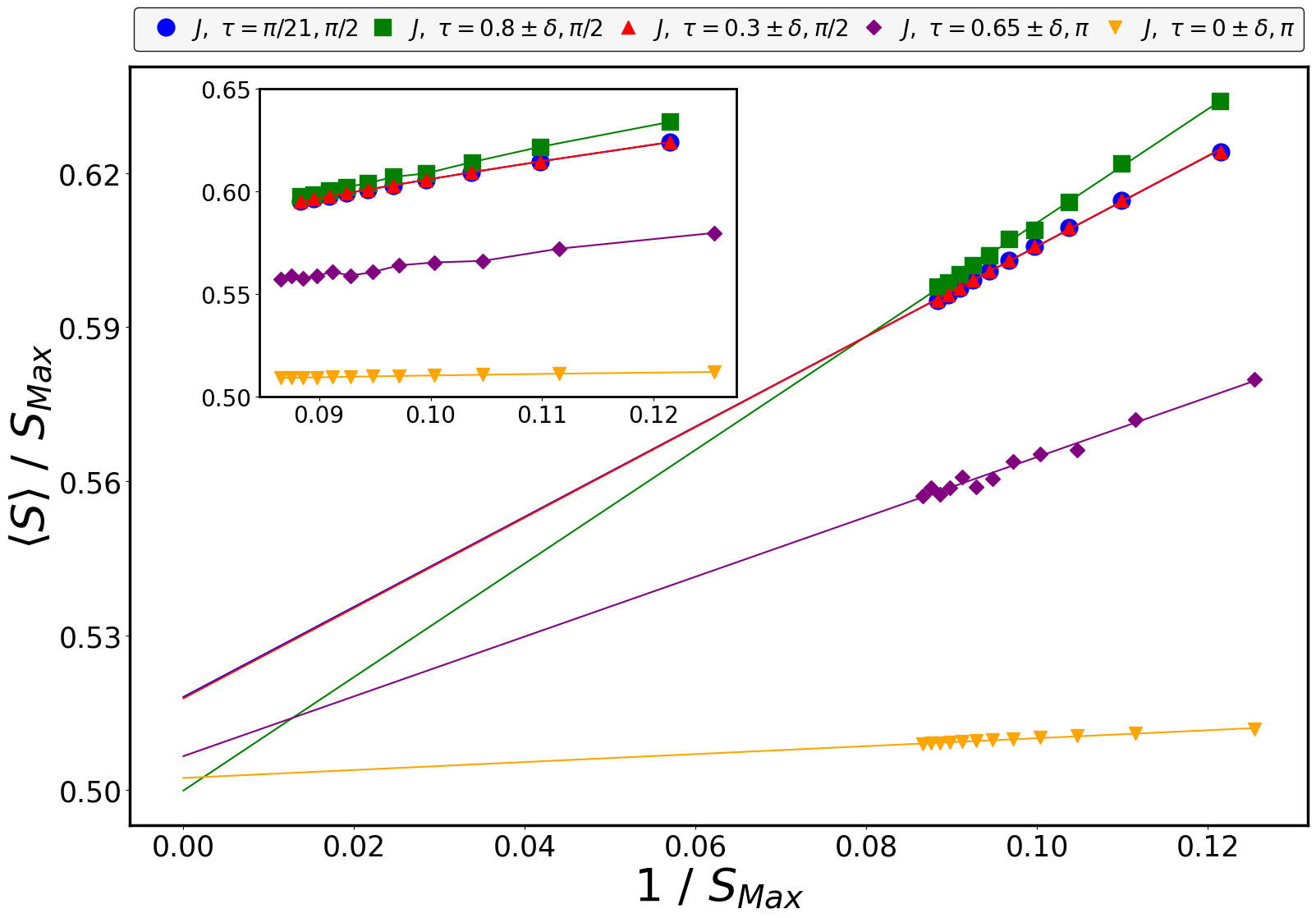}
\caption{The ratio $\langle S \rangle/S_{Max}$ at half-bipartition as a function of the inverse maximal entanglement, $1/S_{Max}=1/\log_2(N/2+1)$. Here we set $\delta=10^{-10}$.}
\label{fig:spectrum48}
\end{figure}

The maximum possible EE for a bipartite system with Hilbert space dimensions $d_A \leq d_B$ is given by $\log_2 d_A$. For a system of $N$ spin-$\frac{1}{2}$ particles restricted to the fully symmetric subspace,, the Hilbert space dimension is reduced from $2^N$ to $N+1$. Consequently, the maximal EE for a bipartition $A: B$ is $\log_2(N_A + 1)$, where $N_A $ denotes the number of particles in subsystem $A$ \cite{kumari2022eigenstate}. We compute the $\langle S \rangle/S_{Max}$ for half bipartition with $1/S_{\mbox{Max}}=1/\log_2(N_A + 1)$ for various parameter sets shown in Fig. \ref{fig:spectrum48}. We  observe that for the rational values of $J$ and  $k'=J\pi j^{p-1}$, even $p$ and $\alpha=m\pi$, where $m$ is an integer, the eigenvalues of the Floquet operator shows high degeneracy in the eigenvalue spectrum. In contrast, the eigenvalues are not degenerated for irrational $J$ and genric values of $k'$. It is known that due to degenerate spectra, the corresponding eigenvectors can get superposed, resulting in increase or decrease in the eigenstate EE. To  address this issue, we lift the degeneracy by applying an infinitesimally small perturbation  $\delta=10^{-10}$ in either of the parameters $J$ and $\alpha$, i.e., $(J\pm \delta, \alpha)$ or $(J,\alpha\pm \delta)$.

We observe   that for any  $k'$, even-$p$ and  $\alpha=m\pi$, the  ratio $\langle S \rangle/S_{Max}$ decreases with $N$.  Using finite-size scaling, we  have demonstrated that the ratio $\langle S \rangle/S_{Max}$  approach to a value that remains very far  below from $1$ in the limit $N\rightarrow \infty$. The limiting value depends on the parameter $k'$, even-$p$ and $\alpha$. These observations are illustrated in Fig. \ref{fig:spectrum48}. We have shown that for any  $k'$, odd-$p$, and  $\alpha=m\pi$, the system shows periodic behavior. Upon introducing a small perturbation
$\delta$, as mentioned earlier, the system shows behavior similar to that observed for even-$p$. Since we observe that for mentioned value of parameters, the asymptotic value of the ratio $\langle S \rangle/S_{Max}$ is less than $1$, this implies QI in our model. Thus, based on all these kinamatic signatures, we conclude that our model is QI for any $k'$, any $p$, and
$\alpha=m\pi$.

\section{Conclusions}\label{sec:example-section6}
In this work, we have studied the Floquet dynamics of the $p$-spin Ising model with infinite-range interactions. This model generalizes the kicked top, which corresponds to the special case $p=2$. We have analytically calculated the unitary operator, its time evolution, the single-qubit RDM, and the entanglement measures for the parameters $\alpha=\pi$, any $k'$ and $p$. We use linear entropy and EE to quantify entanglement in our model for arbitrary initial states and any $N$. We have numerically established QI in our model. It can be  identified through  signatures, like the periodicity of the
entanglement dynamics, periodicity of the time evolution operator,  degenerated spectra, spectral statistics of eigenvalues, the average adjacent gap ratio, SFF and EEE.

Our results show that the  QI in our model depend on the choice of parameters ($k',\alpha$, $p$). From the analytical results, we observe that  for odd-$p$, the entanglement measures and the unitary operator are periodic for any $k'$ and $N$ with period $2$ and $4$ respectively. In this case, the spectrum is  found to be highly degenerate. In contrast, for even-$p$, the entanglement dynamics exhibit periodicity for the rational values of  $k'$ such that $k'=\frac{r}{h}~j^{p-1}\pi$, for  even-$N$ (odd-$N$), with a period of $ph$ ($2^{p-1}h$). In this parameter regime, the model does not possess a well-defined classical limit. Additionally, the time evolution of the unitary operator is periodic, with periods that depend only on $h$, $p$, and $N$ for the same set of parameters.  However, these signatures and spectral degeneracy are absent when $k'$ is
not a rational multiple of $\pi$ and even-$p$.

In such cases where the aforementioned signatures are absent, a thorough analysis of other signatures such as higher-order spectral statistics, higher-order average gap ratio, SFF, and average EEE  is necessary to understand the true nature of the system.  We observe that as the $p$ increases, the eigenphase distribution shows pronounced spikes near $2\pi$, indicating strong eigenphase accumulation in these regions,  arising from quasi-degeneracies. Therefore, we separate the spike region from the spectrum and analyze it independently, where the DOS is nearly uniform. We find that it follows Poisson statistics, indicating the integrable nature of the system.
In contrast, the spike regions exhibit anomalously large values of the average adjacent gap ratio approaching  $0.9$. The SFF (without unfolding) in this region shows a power-law decay without a ramp. These signatures imply that the system is QI.

Apart from spectral diagnostics, we observe that the ratio of the EEE to its maximum value remains significantly below unity in the thermodynamic limit, which is a signature of integrability. After introducing a small perturbation in $k'$ in the range $10^{-4}$ to $10^{-8}$, which lifts the degeneracies, the system  shows Poisson spectral statistics and $\langle S \rangle/S_{Max}$ is far below $1$. These results strongly  shows QI behavior for all $k'$, $N$, $p$ and $\alpha = m\pi$. Based on these numerical and analytical results, we claim that the $p$-spin  model is QI for the said parameters.

We have provided analytical calculations and numerical results for any $N, k', p$, and $\alpha=\pi$ and arbitrary initial state. These results are useful for experimentalists, enabling them to choose the desired initial states,  kicking strength $k'$, and $p$ based on requirements. Furthermore, our findings broaden the range of applicability and provide a  understanding of the system behavior with said parameters and various initial states.  Notably, when the parameter $k'=N^{p-1}J$, the corresponding classical limit no longer exists. Our results could be experimentally verified in platforms such as nuclear magnetic resonance (NMR), ion traps, Floquet synthetic lattices, superconducting qubits, and laser-cooled atoms, trapped-ion quantum computer \cite{neill2016ergodic,Chaudhury09,Krithika2019,meier2019exploring,monroe2021programmable,defenu2023long,ransford202698}. The $p$-spin model serves as a minimal model for adiabatic quantum computation and is closely connected to QAOA \cite{zhou2020quantum}. Moreover, these models can be implemented in analog quantum simulators, including programmable quantum devices, where the interplay between hardware imperfections, simulation errors, and intrinsic nonlinear dynamics remains an open and promising direction for future study. The parameter-dependent dynamics of the model suggest that it could serve as a useful testbed for quantum reservoir computing and quantum extreme learning machines (QELM), enabling the exploration of performance across different regimes by tuning the system parameters \cite{kobayashi2026edge,solanki2026harnessing,xiong2025fundamental}.
\section{Acknowledgment}
The authors express their gratitude to  the Department of Science and Technology (DST) for financial support through project SR/FST/PSI/2017/5(C), sanctioned to the Department of Physics, VNIT Nagpur.
\appendix
\renewcommand\thesection{Appendix \Alph{section}}
\makeatletter
\renewcommand{\@seccntformat}[1]{\csname the#1\endcsname:\quad}
\makeatother

\renewcommand\theequation{\Alph{section}\arabic{equation}}
\renewcommand\thesubsection{\thesection.\arabic{subsection}}
\setcounter{equation}{0}

\section{Analytical Calculations of Entanglement dynamics for the special initial states and even-$N$} \label{appendix:A}
 In this appendix, we  focus on two specific initial states: $\ket{0,0}$, which lies on a period-$4$ orbit, and $\ket{\pi/2,-\pi/2}$, corresponding to a trivial fixed point in the classical phase space. The analytical calculations for the special initial states follow the procedure described in Sec. \ref{sec:example-section46} A for even-$p$.

\subsubsection*{The initial state $\ket{0,0}$}
Let us consider  the spin-coherent state $\ket{0,0}$, represented by $\otimes^N \ket{0}$. The time evolved state can be obtained by the $n$th implementation of the unitary operator $\mathcal{U}$ on the initial state, given as,
\begin{eqnarray}
 \ket{\psi_n}=\mathcal{U}^n \otimes^N\ket{0}=\left(B^+_0\ket{\phi_0^+}+B^-_0\ket{\phi_0^-}\right)/\sqrt{2},
\end{eqnarray}
where the coefficients $A^+_0$, $A^-_0$, $f^n_q$ and $B^{\pm}_0$   can be obtained using Eqs. (\ref{Eq:per2}), (\ref{Eq:per1}), (\ref{Eq:per23}) and (\ref{Eq:11}). The single-qubit RDM is then given as follows:
\begin{equation}
\rho_1(n)=\frac{1}{2}\left(
\begin{array}{cc}
1+ x_n & 0 \\
0 & 1-x_n \\
\end{array}
\right),
\end{equation}
where the coefficients $x_n$ can be written as,
\begin{equation}
 x_n=B^+_{0} (B^-_{0})^*+B^-_{0} (B^+_{0})^*=(-1)^n.
\end{equation}
The linear entropy of single qubit $\rho_1(n)$, is then found to be,
\begin{equation}\label{Eq:entropy21}
 S_{(0,0)}^{(N)}(n,k',p)=\left(1-x_n^2\right)\Big{/}2=0.
\end{equation}
From  Eq. (\ref{Eq:entropy21}), we observe that there is no entanglement between the subsystems for the initial coherent state $\ket{0,0}$ for any even-$N$, any $k'$, even-$p$ and any $n$. This indicates that the system is always in a separable pure state. Although the calculation is shown only for $\ket{0,0}$ any even-$N$, $k'$ and even-$p$, but the states $\ket{0,\phi_0}$ and $\ket{\pi,\mp\pi}$ also give the same results. Similar behavior is observed for these states, odd-$p$ and even-$N$.
\subsubsection*{The initial state $\ket{\pi/2,-\pi/2}$}
Now, we study  another special initial state $\ket{\pi/2,-\pi/2}$, represented by $\otimes^N\ket{+}$, where $\ket{+}=\frac{1}{\sqrt{2}} (\ket{0}+i\ket{1})$. The evolution of this state lies entirely in the positive-party sector.  The time-evolved state can be obtained by the $n$ successive applications of a unitary operator on the initial state, given as,
\begin{eqnarray}\nonumber
 \ket{\psi_n}&=&\mathcal{U}^n \ket{\psi}=\sum_{q=0}^{N/2-1}\left(B^+_q\ket{\phi_{q}^+}\right)+B^+_\frac{N}{2}\ket{\phi_{\frac{N}{2}}^+},
\end{eqnarray}
where the coefficients $A^+_q$, $A^+_{{N}/{2}}$, $B^+_q$ and $B^+_{{N}/{2}}$ can be obtained using Eqs. (\ref{Eq:per2})  and (\ref{Eq:11}).
The single-qubit RDM is then given as follows:
\begin{equation}
\rho_1(n)=\frac{1}{2}\left(
\begin{array}{cc}
 1& \bar{x}_n \\
\bar{x}_n^* & 1 \\
\end{array}
\right),
\end{equation}
where the coefficient  $\bar{x}_n$ is given as follows:
\begin{eqnarray} \nonumber
 \bar{x}_n&=&\sum_{q=1}^{N/2-1}\left\lbrace{\binom{N-1}{q}}\Big{/}\left[\sqrt{{\binom{N}{q}}{\binom{N}{q+1}}}\right]\right\rbrace\left[B^+_{q} (B^+_{q+1})^*-\right.\\ \nonumber && \left.B^+_{q+1}(B^+_{q})^*\right]+\left\lbrace\sqrt{2}~{\binom{N-1}{\frac{N-2}{2}}}\Big{/}\left[\sqrt{{\binom{N}{\frac{N-2}{2}}}{\binom{N}{N/2}}}\right]\right\rbrace\\ && \left[B^+_{\frac{N-2}{2}}\left(B^+_{{N}/{2}}\right)^*-\left( B^-_{\frac{N-2}{2}}\right)^*B^+_{{N}/{2}}\right].
\end{eqnarray}
The eigenvalues of $\rho_1(n)$ are $ \left(1\pm|\bar{x}_n|\right)/2$. The linear entropy of  single-qubit RDM is then given as follows:
\begin{equation}\label{EQ:564566}
 S_{(\pi/2,-\pi/2)}^{(N)}(n,k',p)=\left(1-|{\bar{x}_n}|^2\right)/2.
\end{equation}
 The Eq.~\ref{EQ:564566} allows us to calculate the linear entropy for any $k'$. The resulting dynamics exhibit periodic behavior, with a period similar to that observed for the initial state $\ket{\pi/4,-\pi/4}$, as shown in Figs.~2 and 3.
\section{Analytical Calculations of Entanglement dynamics for the special initial states and odd-$N$} \label{appendix:B}
 In this appendix, we have given detailed analytical calculations for the special initial states using the procedure described  in Sec.~\ref{sec:example-section46} for even-$p$.
\subsubsection*{The initial state $\ket{0,0}$}
 The state $\ket{\psi_n}$ can be obtain by the $n$th implementations  of the unitary operator $\mathcal{U}$ on the initial state, given as,
\begin{eqnarray}\nonumber
 \ket{\psi_n}=\mathcal{U}^n \otimes^N\ket{0}=\left(B^+_0\ket{\phi_0^+}+B^-_0\ket{\phi_0^-}\right)/\sqrt{2},
\end{eqnarray}
where the coefficients $A^+_0$, $A^-_0$, $f^n_q$ and $B^{\pm}_0$   can be obtained using Eqs. (\ref{EQ:Entropy26}), (\ref{EQ:Entropy56}), (\ref{Eq:per23}) and (\ref{Eq:12}). The single-qubit RDM is then given as follows:
\begin{equation}
\rho_1(n)=\frac{1}{2}\left(
\begin{array}{cc}
1+ x_n & 0 \\
0 & 1-x_n \\
\end{array}
\right),
\end{equation}
where the coefficients $x_n$ can be written as,
\begin{equation}
 x_n=B^+_{0} (B^-_{0})^*+B^-_{0} (B^+_{0})^*=(-1)^n
\end{equation}
The linear entropy of a single qubit is then  given as follows:
\begin{equation}\label{Eq:entrop9}
 S_{(0,0)}^{(N)}(n,k',p)=\left(1-x_n^2\right)\Big{/}2=0.
\end{equation}
From Eq. (\ref{Eq:entrop9}), we deduce that the initial coherent state $\ket{0,0}$ exhibits no entanglement between the subsystems for any odd-$N$, even-$p$, any $k'$ and any $n$, indicating that the system is always in a separable pure state.  Although the calculation is shown only for $\ket{0,0}$, but the states $\ket{0,\phi_0}$ and $\ket{\pi,\mp\pi}$ also show the same results. Similar behavior is observed for these states, when  $p$ and $N$ both are odd.
\subsubsection*{The initial state $\ket{\pi/2,-\pi/2}$}
The state $\ket{\psi_n}$ can be obtained by the $n$ successive application of the unitary operator $\mathcal{U}$ on the initial state, expressed as,
\begin{eqnarray}\nonumber
 \ket{\psi_n}&=&\mathcal{U}^n \ket{\psi}=\sum_{q=0}^{(N-1)/2}\left(B^+_q\ket{\phi_{q}^+}\right),
\end{eqnarray}
where the coefficients $A^+_q$ and $B^+_q$ can be obtained using Eqs. (\ref{EQ:Entropy26}) and (\ref{Eq:12}). The single-qubit RDM is then given as follows:
\begin{equation}
\rho_1(n)=\frac{1}{2}\left(
\begin{array}{cc}
 1& \bar{y}_n \\
\bar{y}_n^* & 1 \\
\end{array}
\right),
\end{equation}
where the coefficient  $\bar{y}_n$ is given as follows:
\begin{eqnarray} \nonumber
 \bar{y}_n&=&\sum_{q=1}^{(N-1)/2}\left\lbrace{\binom{N-1}{q}}\Big{/}\left[\sqrt{{\binom{N}{q}}{\binom{N}{q+1}}}\right]\right\rbrace\left[B^+_{q} (B^+_{q+1})^*\right.\\ \nonumber && -\left.B^+_{q+1}(B^+_{q})^*\right].
\end{eqnarray}
The eigenvalues of $\rho_1(n)$ are $ \left(1\pm|\bar{y}_n|\right)/2$. The linear entropy of single qubit $\rho_1(n)$ is then given as follows:
\begin{equation}
 S_{(\pi/2,-\pi/2)}^{(N)}(n,k',p)=\left(1-|{\bar{y}_n}|^2\right)/2.
\end{equation}
For the initial state $\ket{\pi/2, -\pi/2}$, we observe a similar pattern as mentioned earlier for even-$N$. Following the same procedure,  for the case when both $p$ and $N$ are odd, we observe similar nature of entanglement dynamics. The corresponding coefficients and periodicities  are  different from those obtained for even-$N$.

\let\oldaddcontentsline\addcontentsline
\renewcommand{\addcontentsline}[3]{}
\bibliography{refrence13}
\let\addcontentsline\oldaddcontentsline
\end{document}